\documentclass[a4paper, twocolumn,10pt]{quantumarticle}
\pdfoutput=1
\usepackage[utf8]{inputenc}
\usepackage[english]{babel}
\usepackage[T1]{fontenc}
\usepackage{hyperref}
\usepackage{microtype}

\usepackage{amsmath}
\usepackage{amsthm}
\usepackage{mathtools}
\usepackage{diffcoeff}%
\usepackage{braket}%
\usepackage{bm}%
\usepackage[caption=false]{subfig}
\usepackage{adjustbox}

\usepackage{tikz}
\usepackage{quantikz}
\usepackage[compat=0.6]{yquant} %
\useyquantlanguage{groups}
\usetikzlibrary{fit, quotes, calc}
\usetikzlibrary{shapes.geometric, arrows, positioning}

\newtheorem{definition}{Definition}
\newtheorem{proposition}{Proposition}
\usepackage[numbers]{natbib}

\renewcommand{\set}[1]{\{#1\}}
\DeclarePairedDelimiter{\abs}{\lvert}{\rvert}
\DeclarePairedDelimiterX{\norm}[1]{\lVert}{\rVert}{#1}

\newcommand{\Bpar}[1]{\Big(#1\Big)}

\DeclareMathOperator{\Trace}{Tr}

\newcommand{\ketbra}[1]{\ket{#1}\hspace{-0.2em}\bra{#1}}
\newcommand{\Fbar}{\overline{F}}
\newcommand{\probeset}{\mathcal{A}}
\newcommand{\hilb}{\mathcal{H}}
\newcommand{\setDens}{\mathcal{D}(\hilb_n)}

\newcommand{\Wnorm}[1]{\norm{#1}_{W_1}
\newcommand{\empCEM}[1]{\widetilde{C}_\text{EM}\left(#1\right)}
\newcommand{\CEMp}[1]{\empCEM{#1}}
\newcommand{\BempCEM}[1]{\widetilde{C}_\text{EM}\Big(#1\Big)}
\newcommand{\bempCEM}[1]{\widetilde{C}_\text{EM}\big(#1\big)}}

\newcommand{\CHST}{C_\text{HST}}

\newcommand{\CLET}{C_\text{LET}}

\makeatletter
\providecommand*{\input@path}{}
\g@addto@macro\input@path{{./figures//}{../figures//}}
\makeatother

\usepackage{subfiles} %

\begin{document}

    \title{Quantum Wasserstein Compilation:  Unitary Compilation using the Quantum Earth Mover's Distance}
    
    \author{Marvin Richter} 
        \affiliation{Fraunhofer IIS, Fraunhofer Institute for Integrated Circuits IIS, Nuremberg, Germany}%
        \affiliation{Department of Microtechnology and Nanoscience, Chalmers University of Technology, 412 96 Gothenburg, Sweden}
    \author{Abhishek Y. Dubey}%
    \author{Axel Plinge}%
    \author{Christopher Mutschler}%
    \author{Daniel D. Scherer}%
        \affiliation{Fraunhofer IIS, Fraunhofer Institute for Integrated Circuits IIS, Nuremberg, Germany}%
    
    \author{Michael J. Hartmann}
        \affiliation{Friedrich-Alexander University Erlangen-Nürnberg (FAU), Department of Physics, Erlangen, Germany}%
    \date{February 2025}
    
    \maketitle
    
    \begin{abstract}
        Despite advances in the development of quantum computers, the practical application of quantum algorithms requiring deep circuit depths or high-fidelity transformations remains outside the current range of the so-called noisy intermediate-scale quantum devices. Now and beyond, quantum circuit compilation (QCC) is a crucial component of any quantum algorithm execution. Besides translating a circuit into hardware-specific gates, it can optimize circuit depth and adapt to noise. Variational quantum circuit compilation (VQCC) optimizes the parameters of an ansatz according to the goal of reproducing a given unitary transformation. In this work, we present a VQCC-objective function called the quantum Wasserstein compilation (QWC) cost function based on the quantum Wasserstein distance of order~1. We show that the QWC cost function upper bounds the average infidelity of two circuits. An estimation method based on measurements of local Pauli-observable is utilized in a generative adversarial network to learn a given quantum circuit. We demonstrate the efficacy of the QWC cost function by compiling hardware efficient ansatz (HEA) as both the target and the ansatz and comparing to cost functions such as the Loschmidt echo test (LET) and the Hilbert-Schmidt test (HST). Finally, our experiments demonstrate that QWC as a cost function is the least affected by barren plateaus when compared to LET and HST for deep enough circuits.
    \end{abstract}

\section{Introduction}
    \label{sec:intro}
    The compilation of quantum circuits is as crucial to quantum computing as the compilation of human-readable code into executable machine language is to traditional computing. By compilation, we are able to focus on the fundamental operations in both quantum and traditional computing thanks to the abstraction of the underlying complexity. 
    
    Quantum circuit compilation (QCC) entails translating a target quantum algorithm into an executable quantum circuit compatible with real quantum computing hardware. This intricate process must account for the target hardware constraints, including the available gate alphabet, qubit connection graph, and depth restrictions. Additionally, a strategic approach may consider individual error rates of single and two-qubit operations, single-qubit decoherence rates, and readout errors during the rewriting process to minimize the probability of errors during execution. In the context of noisy intermediate-scale quantum (NISQ) computing, these optimizations are not mere conveniences, but pivotal elements~\cite{preskillQuantumComputingNISQ2018}. 
    The considerations in the QCC process thus underscore its critical importance in the era of NISQ computing.
    
    One approach to QCC is based on the variational quantum computing paradigm, which focuses on optimizing the parameters of a circuit to minimize a cost function. Several cost functions have been developed for this purpose, starting with the work of Khatri et al.~\cite{khatriQuantumassistedQuantumCompiling2019}, where the similarity between the target unitary and the ansatz was evaluated directly on the quantum computer. This method allows for bypassing the need for exponentially many resources that arise from the increasing complexity of the Hilbert space of quantum states.
    Recent findings indicate that current methods of variational quantum circuit compilation (VQCC) do not fully exploit the potential of the data that is made available to them, because their data requirements grow exponentially with the size of the target system \cite{cincioLearningQuantumAlgorithm2018,cincioMachineLearningNoiseResilient2021}. However, based on the findings of Caro et al.~\cite{caroGeneralizationQuantumMachine2022}, a polynomial amount of training data should be sufficient to approximately compile a target circuit, when a loss function based on the expectation value of an observable is used. This encourages us to look for improved methods of VQCC.
    
    Until now, methods of VQCC have been closely related to the overlap of quantum states. However, the state overlap has two fundamental properties, making it an ineffective cost function. Firstly, certain parts of the system can completely dominate the state overlap. For instance, if the state of a subsystem is orthogonal to the state of its variational counterpart, the overlap between the overall system states becomes zero, in addition to the overlap between the subsystem states. Secondly, the state overlap for two randomly picked quantum states decreases exponentially with system size. The vanishing of the state overlap also results in a learning signal that is exponentially smaller and hence exponentially more expensive to measure when we use the state overlap as an objective function.

\subsection{Contributions}
    The scope of our work is defined as follows: we focus on continuous parameter optimization rather than circuit structure learning, and our numerical experiments use the hardware-efficient ansatz (HEA) for both target and ansatz circuits. All simulations are limited to noiseless environments with up to 8 qubits.
    Our work makes the following contributions to the field of VQCC:

    \begin{itemize}
        \item Introduction of the Quantum Wasserstein Cost (QWC) Function for Unitary Compilation: We propose a novel cost function based on the quantum Wasserstein distance of order 1. Unlike traditional unitarily invariant metrics, this distance (also called Earth Mover's distance) provides an alternative approach to measuring distances between quantum states. It grows linearly with system size and is additive~\cite{kianiLearningQuantumData2022} rather than multiplicative for subsystems, preventing any subsystem from dominating the overall distance.
        \item Theoretical motivation: We theoretically motivate the usage of QWC which extends the quantum $W_1$ distance to compare unitary operations (see Section~\ref{sec:qwc-ideal-cost}). This approach is based on simultaneously reducing the estimated $W_1$ distance between the output states across multiple input states. We prove that QWC provides an upper bound for the average infidelity between unitary transformations, establishing its validity for circuit compilation tasks. Moreover, our approach differs from that presented in Ref.~\cite{qiuQuantumWassersteinDistance2024} on unitary compilation, as our distance lower bounds the distance introduced therein. 
        \item GAN-inspired architecture: Our implementation combines quantum-state discrimination with generative adversarial networks. The method comprises two key components: the generator, consisting of the ansatz, and the discriminator, measuring the empirical cost function based on the averaged Wasserstein distance between the states generated by the target and the ansatz. We make our complete implementation available as an open-source GitHub repository~\cite{Richter_quantum-wasserstein-compilation_2025}.
        \item Analysis of Barren Plateaus: Through numerical experiments, we demonstrate that the one-step gradients of our cost function are least affected by barren plateaus as we scale to larger qubit numbers and deeper circuits. This avoids one of the key challenges in variational quantum algorithms, potentially enabling more effective training for larger quantum systems.
    \end{itemize}

    The paper is organized as follows: Section~\ref{sec:fundamentals} introduces the preliminaries of unitary compilation along with the various cost functions used in the literature. Section~\ref{sec:related_work} reviews previous work on variational compilation methods. Section~\ref{sec:our_work} discusses the concepts which are important in our approach. Section~\ref{sec:exps} details the experimental setup and discusses the results. Section~\ref{sec:conclusion} concludes the paper with a discussion of our approach. The Appendix provides a brief overview of the theoretical background.

\section{Preliminaries}
    \label{sec:fundamentals}
    \subsection{Unitary Compilation}
        \label{sec:compilation}
\begin{figure}
            \centering
            \includegraphics[width=0.9\linewidth]{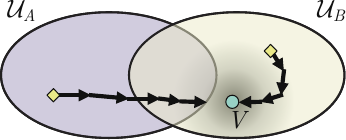}
            \caption{The two manifolds $\mathcal{U}_A$ and $\mathcal{U}_B$ represent two families of unitaries created by different ansätze and $\diamond$ denotes the starting point of the optimization of the continuous parameters. Here, the ansatz $B$ can reach the optimal unitary~$V$. In contrast, ansatz $A$ only admits an (possibly bad) approximation. Note: The optimization landscape is non-convex.}
            \label{fig:ansatz}
        \end{figure}

        In this section, we will review unitary compilation in the variational quantum machine learning framework \cite{cerezoVariationalQuantumAlgorithms2021}. Here, compilation describes the process of finding a decomposition of a unitary transformation $V$ into a specific set of parameterized unitaries available on the hardware $\set{U_i (\theta_i)}$, i.e.
        \begin{align}
            V \approx U_1(\theta_1) U_2(\theta_2) U_3(\theta_3) \dots U_P(\theta_P) \eqqcolon U(\bm \theta)\;,
        \end{align}
        with $P$ parameters $\theta_i$. The unitary compilation process consists of two steps: (a) choose an appropriate ansatz represented by the sequence and types of parameterized unitaries $U_i$, and (b) determine the optimal parameters (see Fig.~\ref{fig:ansatz}).
        
        Choosing an appropriate \textit{ad hoc} ansatz presents a significant challenge due to the fundamental trade-off between ansatz expressivity and trainability. Higher expressivity is linked to vanishing gradients \cite[]{holmesConnectingAnsatzExpressibility2022}. Therefore, the selection of an ansatz demands the use of intuition and the application of prior knowledge about the target unitary. The underlying symmetries might be used to pick an ansatz that is not excessively expressive, but still includes an optimal solution \cite[]{meleAvoidingBarrenPlateaus2022}.
        
        Addressing the issue of expressivity versus trainability necessitates exploring strategies to update the structure. One possible approach includes adding layers incrementally to the ansatz until a satisfactory approximation of the target unitary is achieved \cite{khatriQuantumassistedQuantumCompiling2019}. This method offers the advantage of progressively enhancing the ansatz's expressivity. During the extension, the complexity increase can be limited by only accepting updates that improve the approximation quality.
        
        Another approach to increasing the expressivity of an ansatz, while maintaining control over its complexity, involves a technique called variable ansatz \cite[]{bilkisSemiagnosticAnsatzVariable2023}. This optimization technique adds and removes gate sequences interleaved with the continuous parameter optimization. This enables searching for appropriate solutions while keeping the candidates shallow and thus potentially trainable for local cost functions \cite{cerezoCostFunctionDependent2021}.
        
        The technique that we developed in this work tackles the problem of finding optimal parameters for a given ansatz. In other words, we train a parameterized quantum circuit, represented by the unitary operator $U(\bm \theta)$, such that it is close to a given target unitary operator $V$. Since closeness for unitary transformations can be defined in several ways, various application-tailored distance measures have been defined.
        
         Unitary compilation can be classified into three categories: (a) full unitary matrix compilation (FUMC), (b) fixed input states compilation (FISC), and (c)~single input state compilation (SISC). For example, FISC can be used in classical-into-quantum data encoding where the set of input states is limited. SISC finds application in state preparation circuits for quantum chemistry. 
        
        In FUMC, the goal is to reproduce the complete unitary matrix and hence mimic the target evolution of every possible input state. In consequence, the average fidelity is the natural figure of merit for FUMC.
        \begin{definition}[Average Fidelity \cite{nielsenSimpleFormulaAverage2002, khatriQuantumassistedQuantumCompiling2019}]
        Given two unitary transformations $U$ and $V$, the average fidelity between them is defined as:
        \begin{align}
            \label{eq:haar-av-fid}
            \Fbar(U, V) = \int \dl \psi \abs{\braket{\psi | V^\dagger U | \psi}}^2\;.
        \end{align}
        Here, $\dl \psi$ represents the integration over the unitarily invariant Fubini-Study measure on pure states.
        \end{definition}
        
        The average fidelity quantifies how closely the two transformations resemble each other for arbitrary input states.
        Alternatively in FISC, for cases where we only aim to reproduce the evolution of a fixed set $\probeset$ of quantum states, we can use a simpler figure of merit---the set-average state fidelity, defined as:
        \begin{align}
            \label{eq:set-avg-fidelity}
            F(U,V, \probeset) = \frac{1}{|\probeset|} \sum_{\ket{\psi}\in\probeset}\abs{\braket{\psi| V^{\dagger}U| \psi}}^2\;,
        \end{align}
        where $|\probeset|$ denotes the cardinality of set $\probeset$. In SISC, we only consider a single state, that is, $\mathcal{A}=1$.
        
    \subsection{Cost Functions of Variational Compilation}
    
        The variational compilation process optimizes a parameterized unitary operator $U(\bm \theta)$ to approximate a target unitary $V$ by minimizing specific cost functions. We introduce two metrics for this optimization that we will use as a comparison to our own metric. The first, the Hilbert-Schmidt test, was proposed by Khatri et al.~\cite{khatriQuantumassistedQuantumCompiling2019} for VQCC and can be implemented on a quantum computer using Bell states and Bell measurements when both unitaries are coherently accessible (i.e., on the same quantum hardware or in an entangled system). For $n$ qubits, this metric is defined as:
        \begin{equation}
            \label{eq:HST}
            C_{\text{HST}} = 1-\frac{|\text{Tr}(V^{\dagger}U)|^2}{4^n}\;.
        \end{equation}
        Notice that this metric does not depend on a set of input states and is used for FUMC. 
        Minimization of this cost function ensures the closeness between the unitary $U$ and $V$ since it is related to the average fidelity defined in Eq.~(\ref{eq:haar-av-fid}) by the relation
        \begin{equation}
            \label{eq:HST-Fbar_relatn}
            \bar{F}(U, V) = \frac{2^n+|\text{Tr}(V^{\dagger}U)|^2}{4^n + 2^n} \;.
        \end{equation}
        
        The second cost function that we will use as a comparison is based on the Loschmidt echo~\cite[]{goussevLoschmidtEcho2012} and was used as the Loschmidt echo test (LET) for SISC in~\cite[]{sharmaNoiseResilienceVariational2020}. 
        The Loschmidt echo quantifies the overlap of an initial state~$|\psi_0\rangle$ and the evolution of the same state under unitary $V^{\dagger}U$. 
        For a fixed input state~$|\psi_0 \rangle$, this overlap defines the LET cost function as $|\langle \psi_0 | V^{\dagger}U| \psi_0 \rangle|^2$. 
        To extend this metric for FUMC, we average over a set of input states:
        \begin{equation}
            \label{eq:LET}
            C_{\text{LET}} = 1 - \frac{1}{|\probeset|}\sum_{\psi \in \probeset}|\langle \psi | V^{\dagger}U| \psi \rangle|^2\;.
        \end{equation}
        
        Both cost functions, $C_{\text{HST}}$ and $C_{\text{LET}}$, are global cost functions and suffer from barren plateaus~\cite{cerezoCostFunctionDependent2021}. To address this, local HST~(LHST) and local LET (LLET) were introduced. The detailed circuit implementations of HST, LET and their local forms are given in the Appendix~\ref{sec:cost-functions-for-vqcc}.
    
    \subsection{The Quantum Wasserstein Distance of Order 1}\label{sec:w1-dist}
        De Palma et al. \cite{depalmaQuantumWassersteinDistance2021} introduce the Wasserstein distance of order 1 for quantum states (or the quantum $W_1$ distance). It is a generalization of the classical Wasserstein distance for probability distributions (also called the earth mover's distance) to quantum states. It has an interpretation as a continuous version of a quantum Hamming distance, which could be intuitively described as the number of differing qubits. In the following, we will reproduce the dual formulation of the quantum $W_1$ distance (which is a semidefinite program) between two quantum states $\rho,\sigma\in\setDens$ where $\setDens$ is the set of density operators.
        
        \begin{proposition}[\textit{De Palma et al.} \cite{depalmaQuantumWassersteinDistance2021}]
            \label{prop:dual-form-EM}
            For two $n$-qubit quantum states $\rho,\sigma\in\setDens$, the quantum $W_1$ distance admits a dual formulation with strong duality,
            \begin{align}
            \label{eqn:dual-EM}
                W_1(\rho,\sigma)&=\norm{\rho-\sigma}_{W_1} \nonumber \\ 
                &=\max (\mathrm{Tr}[H(\rho-\sigma)]:H\in\mathcal{M}_n, ||H||_L\leq1),
            \end{align}
            where $\mathcal{M}_n$ denotes the set of observables on $\hilb_n$ and $||\cdot||_L$ the quantum Lipschitz constant\cite{depalmaQuantumWassersteinDistance2021}.
        \end{proposition}
        
        In the context of VQCC, the quantum $W_1$ distance has several intriguing properties, the most important of which is that it is not unitarily invariant. Although this does not seem like an advantage, it makes the quantum $W_1$ distance fundamentally different from the better known distance measures of quantum states like the trace distance or the quantum fidelity. As Kiani et al. \cite{kianiLearningQuantumData2022} pointed out, this property facilitates the learning of quantum states: Consider wanting to learn and reproduce a state $\ket{\text{GHZ}_2}\ket{1}$ from the initial state $\ket{000}$. If we change to $\ket{\text{GHZ}_2}\ket{0}$ from the initial state during learning, then this significant improvement towards the target should be admitted by the cost function. No unitarily invariant distance can discriminate between the three pairwise orthogonal states, and hence indicate the improvement. 
        
        Furthermore, the quantum $W_1$ distance is superadditive with respect to the tensor product, i.e.,
        $W_1 (\rho, \sigma) \geq W_1(\rho_{1..k}, \sigma_{1...k}) + W_1(\rho_{k+1..n}, \sigma_{k+1...n})$ for two $n$-qubit quantum states $\rho, \sigma$ and any $k = 1,...,n-1$. $\rho_{1..k}$ and $\rho_{k+1...n}$ are the marginal states over the first $k$ and last $n-k$ qubits, respectively. This ensures good linear scaling of the distance measure with the number of qubits and, consequently, for the gradient calculations. 
        
        To justify the usage of the quantum $W_1$ distance in VQCC, we examine the containment given by the trace norm $\norm{\cdot}_1$ \cite{depalmaQuantumWassersteinDistance2021},
        \begin{align}
            \label{eq:W1-Trace-norm}
            \frac{1}{2}\norm{\rho - \sigma}_1 \leq \Wnorm{\rho - \sigma } \leq \frac{n}{2} \norm{\rho - \sigma}_1\;.
        \end{align}
        From there, we can derive (see Appendix~\ref{sec:em-to-fidelity}) an upper bound for the infidelity for small quantum $W_1$ distances of mixed states, i.e. $0 \leq \Wnorm{\rho-\sigma} \leq 1$,
        \begin{align}
           2 \Wnorm{\rho - \sigma} \geq 1 -  F(\rho, \sigma) \;.
        \end{align}
        Additionally, we find that a stronger upper bound (without constraining to the small $W_1$ distance regime) holds w.r.t the infidelity between pure states, 
        \begin{align}
            \label{eq:W1-to-Fidelity}
            \big\lVert\ketbra{\psi}-\ketbra{\phi}\big\lVert_{W_1}^2 \geq  1-F(\ket{\psi},\ket{\phi}) \;.
        \end{align}
        This upper bound for the infidelity of pure states in terms of the quantum $W_1$ norm will motivate our definition of the quantum Wasserstein compilation cost.

\section{Related Work}
    \label{sec:related_work}
    Using variational quantum circuits for quantum compilation was introduced by Khatri et al.~\cite{khatriQuantumassistedQuantumCompiling2019}. They demonstrated successful training of cost functions like HST and LHST for unitaries up to 9 qubits, with and without noise. However, they also showed the presence of barren plateaus in the gradients of these cost functions even with depth-one circuits. Barren plateaus in variational quantum circuits have been theoretically proven to occur when circuit depth scales polynomially, $D \in \mathcal{O}(\mathrm{poly}(n))$, with the number of qubits $n$~\cite{mccleanBarrenPlateausQuantum2018}.
    Building on this, Cerezo et al. \cite{cerezoCostFunctionDependent2021} provided bounds on the variance of gradients for global and local cost functions as a function of circuit depth $D$. So, a key focus has been on addressing the barren plateau problem. One approach was the initialization strategy in Ref.~\cite{grantInitializationStrategyAddressing2019}, which kept the ansatz close to the identity to maintain constant gradient variance scaling. An analytical study of Wasserstein distance between unitaries along with the properties of the distance was also done in Ref.~\cite{qiuQuantumWassersteinDistance2024}, providing a metric for comparing quantum gates.

    Additionally, prior work has looked at the sample complexity for successful learning and generalization in variational quantum algorithms. Caro et al. ~\cite{caroGeneralizationQuantumMachine2022} derived bounds showing the generalization error (the difference between the prediction and training errors) scales approximately as $\sqrt{T/N}$, where $T$ is the number of parametrized gates and $N$ is the size of the training data.

\section{Our Work}
    \label{sec:our_work}
    In this section, we introduce the quantum Wasserstein compilation (QWC) as an extension of the quantum $W_1$ distance for comparing unitaries. It is based on the idea of simultaneously reducing the estimated $W_1$ distance of output states for multiple different input states. In Section~\ref{sec:qwc-ideal-cost}, we derive an ideal cost function as the minimization over the average $W_1$ distance for all pure quantum states. We also indicate its significance for unitary compiling. Then, in Section~\ref{sec:qwc-empirical} we will formulate an approximation of the QWC cost function that is directly accessible by taking the mean over representative set of quantum states and estimating the cost function from measuring Pauli observables. In Section~\ref{sec:qwc-probe-states}, we will briefly describe the representative state ensemble needed as input to the unitaries during compilation. Finally, in Section~\ref{sec:qwc-training}, we will describe the complete learning algorithm. 

    In Section~\ref{sec:exps}, a numerical study follows where we set a focus on determining working regimes for the hyperparameters, namely the $k$-locality of the Pauli observables used for estimation in Section~\ref{sec:hyperparams} and the size of the state ensemble in Section~\ref{sec:data-demand}. After finding the working parameters, we will compare QWC with HST and LET in Section~\ref{sec:results-training}. In particular, we point out its advantage regarding barren plateaus during training in Section~\ref{sec:results-bp}.
    
    \begin{figure*}
        \includegraphics[width=\linewidth]{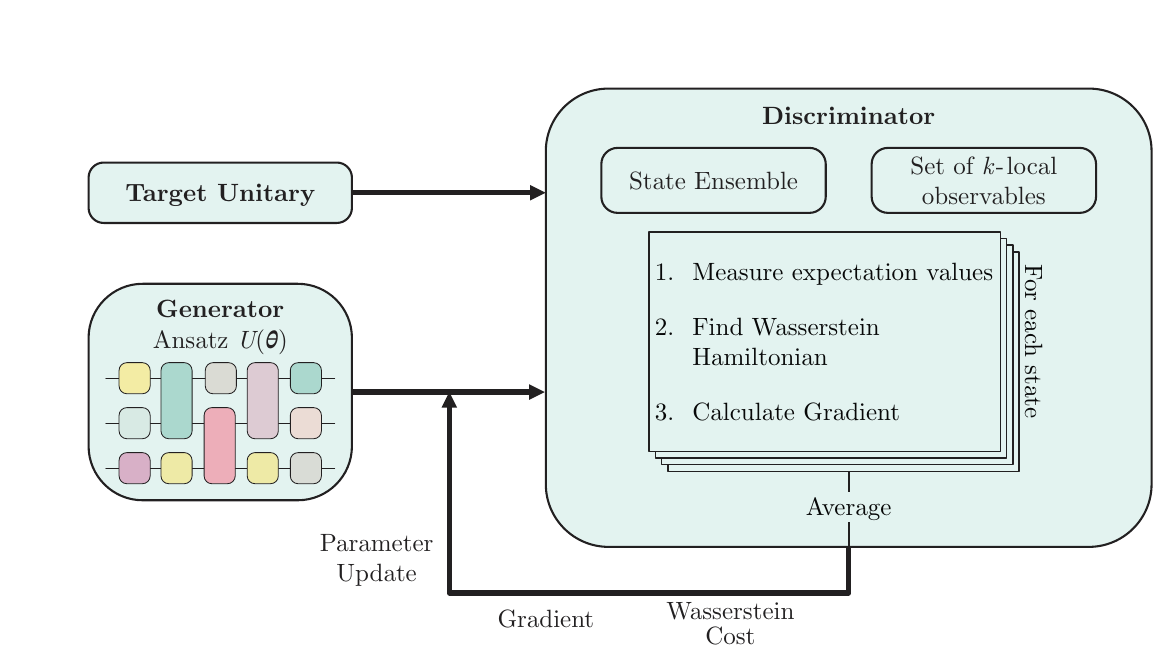}
        \caption{Overview of the compiling algorithm. The target unitary and the parameterized circuit acting as the generator are assessed by the discriminator which calculates the Wasserstein compilation cost. The distance estimation requires a state ensemble acting as input states for target and generator and a set of $k$-local observables whose expectation values are measured. A Wasserstein Hamiltonian can be constructed from the differences of the expectation values and the gradient of the averaged cost can be used for updating the parameters of the generator.}
        \label{fig:alg}
    \end{figure*}
    
    \subsection{Ideal Cost}
        \label{sec:qwc-ideal-cost}
        As outlined in Section~\ref{sec:w1-dist}, the quantum $W_1$ distance is a measure of the closeness of two quantum states.  We will now extend this distance to measuring the closeness of two unitary operators, $U$ and $V$, by applying the operators on (pure) quantum states and measuring the pairwise distances:
        
        \begin{definition}[Quantum Wasserstein Compilation Cost]
            \label{def:qwc}
            Let $U,V$ be unitary operators on $\hilb$ and $\ket{\psi}$ be a quantum state in $\hilb$. Then the quantum Wasserstein compilation cost is defined as
            \begin{align}
                \label{eq:CEM-exact}
                C_{QW}(U,V) = \int_{\psi}\dl{\psi} \: W_1^2\Bpar{U\ket{\psi}, V\ket{\psi}}\;,
            \end{align}
            where $\dl\psi$ is the Fubini-Study metric.
        \end{definition}
        We chose to define the QWC cost in Eq.~(\ref{eq:CEM-exact}) as the squared $W_1$ distance since it then acts directly as an upper bound for the average infidelity as shown below:
        \begin{proposition}
            \label{prop:CEM-FBar}
            Let $U, V$ be unitary operators on $\hilb$. Then the following inequality holds between the QWC cost $C_{QW}(U,V)$ and the average fidelity $\Fbar(U,V)$
            \begin{align}
                \label{eq:CEM-FBar}
                C_{QW}(U,V) \geq 1 - \Fbar(U,V)\;.
            \end{align}
        \end{proposition}
        \begin{proof}
            We use that the quantum $W_1$ norm is an upper bound for the infidelity that we derive in Appendix~\ref{sec:em-to-fidelity}. Starting from the definition of the QWC cost in Eq.~(\ref{eq:CEM-exact}), we can directly upper bound the average fidelity:
            \begin{align}
                C_{QW}(U,V) & =  \int_{\psi} \dl{\psi} \: W_1^2\Bpar{U\ket{\psi}, V\ket{\psi}}  \\
                          & \geq \int_\psi \dl{\psi} \:  \left(1 - F(U\ket{\psi}, V\ket{\psi})\right) \\
                          & = 1 - \Fbar(U,V) \;.
            \end{align}
        \end{proof}
    
        Proposition~\ref{prop:CEM-FBar} provides a theoretical link between $C_\text{QW}$ and the average infidelity. By establishing a direct upper bound on the average infidelity, this result transforms the QWC cost into a meaningful optimization objective for VQCC. During the compilation process, minimizing $C_\text{QW}(U,V)$ directly corresponds to maximizing the fidelity between the parameterized circuit $U(\bm \theta)$ and the target circuit $V$. This means that as the compilation algorithm drives the QWC cost lower, it simultaneously improves the quantum circuit's ability to approximate the target unitary transformation across a diverse set of input states. \newline

    \subsection{Empirical Cost}
        \label{sec:qwc-empirical}
        In order to calculate the cost in Eq.~(\ref{eq:CEM-exact}) we need to first estimate the quantum $W_1$ distance as defined in Eq.~(\ref{eqn:dual-EM}). For this, as proposed by Kiani et al.~\cite{kianiLearningQuantumData2022} we begin by choosing the observables that satisfy the quantum Lipschitz condition. We use the ansatz for $H$, which is a weighted sum of locally acting Pauli observables.
        \begin{align}
            \label{eq:hamiltonian}
            H = \sum_m w_m H_m \quad H_m = \bigotimes_{j=1}^n \sigma_{P_j}^{(j)} \quad P_j \in \{ I,X,Y,Z \}.
        \end{align}
        This ansatz has $4^n$ observables, growing exponentially with the number of qubits. To reduce this growth, we are restricting the set of observables $\mathcal{M}_n$ to $\mathcal{M}_n^{(k)}$~\cite{kianiLearningQuantumData2022}, which is defined as the set of Pauli strings that act non-trivially only on a subset of $k$ qubits, and is referred to as $k$-local Pauli observables. Using local Pauli operators restricts the growth of the number of Pauli observable to $\mathcal{O}(n^k)$ for $k \ll n $. Thus we instead have the approximation
        \begin{align}
            \label{eq:k-local-EM}
            W_1^{(k)} = \max (\text{Tr}[H(\rho - \sigma)]:H\in\mathcal{M}_n^{(k)},||H||_L < 1)\;.
        \end{align}
        
        Moreover, the space of all quantum states is growing exponentially fast in system size and even for small qubit numbers, is inaccessibly large. To overcome this hurdle, we use a \emph{state ensemble} $\probeset = \{|\psi\rangle_s\}$, restrict to $k$-local observables and measure the empirical distance:
        \begin{align}
            \label{eq:empirical_CEM}
            \tilde{C}_{QW}^{(k)}(U,V,\probeset) = \frac{1}{|\probeset|} \sum_{\psi\in\probeset} \left(W_1^{(k)}(U \ket{\psi}, V \ket{\psi})\right)^2\;.
        \end{align}
        
        The choice and size of the state ensemble $\probeset$ are decisive for the practical use of $\tilde{C}_{QW}^{(k)}$ as an optimization objective in VQCC. In the limit of infinitely many states that are sampled according to the Fubini-Study metric and no restriction on the locality of Pauli operators, the empirical quantum Wasserstein compilation distance becomes equivalent to the ideal distance from Eq.~(\ref{eq:CEM-exact}). In contrast to the Wasserstein distance for unitaries defined in Ref.~\cite{qiuQuantumWassersteinDistance2024} which is the maximum distance over all possible states, our cost function naturally acts as a lower bound to their definition. Moreover, they do not provide a method for estimating the distance for arbitrary multi-qubit unitaries.
        
        The derivatives of the cost function $\tilde{C}_{QW}^{(k)}(U,V,\probeset)$ with respect to a parameter $\theta\in\bm \theta$ can be directly calculated from the respective derivative of the $W_1$ distance~\cite{kianiLearningQuantumData2022}, around a value $t$:
    
        \begin{widetext}
            \begin{eqnarray}
                \label{eq:derivatives}
                \diffp*{\tilde{C}_{QW}^{(k)}{U(\theta),V, \probeset}}{\theta}[\theta=t]  =  \frac{1}{|\probeset|} \sum_{\ket{\psi_a}\in\probeset} 2 W_1^{(k)}\Bigl(U(t) \ket{\psi_a}, V \ket{\psi_a}\Bigr) \cdot 
                 \diffp*{W_1^{(k)} \Bigl(U(\theta) \ket{\psi_a}, V \ket{\psi_a}\Bigr)}{\theta}[\theta=t] \;.
            \end{eqnarray}
        \end{widetext}
    
        The derivative $\diffp*{W_1^{(k)} \Bigl(U(\theta) \ket{\psi}, V \ket{\psi}\Bigr)}{\theta}[\theta=t]$ can be evaluated using standard techniques such as the parameter-shift rule \cite{schuldEvaluatingAnalyticGradients2018}. A detailed derivation of these gradients is provided in Appendix I of Ref.~\cite{kianiLearningQuantumData2022}.
        
        Since we now have the cost function and its gradients, the only missing building block for learning unitaries is the choice of the state ensemble.
        
    \subsection{State Ensembles}
        \label{sec:qwc-probe-states}
        Our full unitary matrix compilation method depends on a state ensemble $\probeset$. Caro et al.~\cite{caroOutofdistributionGeneralizationLearning2023} showed that when average infidelity is used as a cost function, learning over a locally scrambled ensemble is equivalent to learning over the uniform distribution of states over the complete Hilbert space. This seminal result paves the way to use an ensemble of product states $\mathcal{S}_{\text{Haar}^{\otimes n}_1}$ where each product state is the combination of Haar-random single-qubit states. Random product states can be prepared using a shallow circuit of depth three in contrast to multi-qubit Haar-random states which require deep circuits.
        
        While the sizes are determined for SISC and FISC, the number of states used to determine the empirical cost function is an important hyperparameter of FUMC. QWC for FUMC can use a fixed set $\probeset$ of input states, which we will call fixed mode, or sample input states in each compilation step, which we call sampling mode.
        
        It is an open question how much data in the form of quantum states is needed to successfully learn a given unitary. Some authors expect that compilation from data requires very large datasets \cite{sharmaReformulationNoFreeLunchTheorem2022,polandNoFreeLunch2020}. Recent results show that it is  sufficient to have training data that has size polynomial in the number of qubits~\cite{caroGeneralizationQuantumMachine2022}. The argument is based on the proposition that the required size of the training data is roughly linear in the number of parameterized gates. As a matter of fact, virtually all the ansätze used in practice have significantly fewer parameters than the degrees of freedom of a corresponding unitary. Furthermore, the parameters are often not independent, leading to a further reduction of the actual number of degrees of freedom.
        
        In this work, we will utilize another approximation: a SU(2) transformation $U_3(\theta,\phi,\lambda)$, parameterized by 3 angles, is applied to each qubit. Sampling each parameter randomly and uniformly between $(-\pi,\pi]$ creates a random product state. It is well known that such a transformation $U_3$ can be decomposed into three rotational gates, for example using Z- and Y-rotations:
        \begin{align}
            \label{eq:u3-gate}
            U_3(\theta,\phi,\lambda) = R_\text{Z}(\lambda)R_\text{Y}(\phi)R_\text{Z}(\theta) \;.
        \end{align}
    
         Using a fixed set of states might decrease the number of circuit evaluations since the Pauli measurements for the state ensemble under the target evolution can be done in advance\footnote{We assume no restrictions on classical memory to store the measurement results. The number of expectation values scales as $\mathcal{O}(M |\probeset|)$ where $M$ denotes the number of Pauli measurements and $|\probeset|$ the number of states}. On the other hand, using a set of states in the sampling mode increases computation since the target unitary needs to be measured for the sampled states. We discuss our choice in Section~\ref{sec:exps}.

    \subsection{Learning a Unitary using QWC}
    
        \label{sec:qwc-training}
        In the previous sections, we introduced the empirical quantum Wasserstein compilation cost and its derivatives for parameterized unitaries (see Eq.~(\ref{eq:empirical_CEM})-(\ref{eq:derivatives})). Based on these ideas, we can formulate a procedure to learn a target unitary $V$ presented in Fig.~\ref{fig:alg}. 

        The compilation is in the form of a quantum Wasserstein Generative Adversarial Net (GAN)~\cite{kianiLearningQuantumData2022}. The generator is a variational quantum circuit with parameters $\bm \theta$ that output a state $G(\bm \theta)$, and the discriminator is the estimator of the averaged $W_1$ distance. Quantum GAN are quantum adversarial games, in which the Nash equilibrium can be reached in an all-quantum game if the generator is expressive enough to reproduce the target and the discriminator has the ability to find a measurement that discriminates them~\cite{lloydQuantumGenerativeAdversarial2018}. The expressivity of a quantum circuit specifies the set of unitary transformations it can reproduce, and, of course, for a successful approximate compilation, there should be an approximation of the target unitary in this set. Due to the limited scope of this study, the expressivity of the generator is not explicitly addressed, and the experiments in Section~\ref{sec:exps} were designed in a way that guaranteed sufficient expressivity of the generator. The discrimination ability, on the other hand, depends on several factors that we examine in this work.
        
        The first step of every optimization is measuring the expectation values of the Pauli observables $H_m \in \mathcal{M}_n^{(k)}$ for every input state $\ket{\psi_a}\in \probeset$ after evolving with the generator ansatz and the target. We denote the evolved set of states as $\{G(\bm \theta)\ket{\psi_a} \}$ (with density matrix $\rho(\bm \theta)$) and $\{V\ket{\psi_a} \}$ (with density matrix $\sigma$). The expectation value difference is given by $c_m = \text{Tr}(\rho(\bm \theta)H_m)-\text{Tr}(\sigma H_m)$. If the states and the observables are fixed, the result of the target can be cached and does not need to be measured again. Then we solve the linear program for the weights $w_m$
        \begin{equation}
            \begin{array}{ll@{}ll}
               \text{maximize}  & \sum_m w_m c_m & \\
               \text{constraint}  & \sum_{m:i\in\mathcal{I}_m} |w_m| \leq 1/2 \quad \forall i \in [n] \;.
            \end{array}
        \end{equation}
        Note that the weights $w_i$ are sparse with only $n$ non-zero entries and the corresponding Pauli operators are called \textit{active}~\cite{kianiLearningQuantumData2022}.
    
        The state-wise quantum $W_1$ distances $W_1^{(k)}$ can be  measured from Eq.~(\ref{eq:k-local-EM}) with the Hamiltonian $H_{W}=\sum_{n\in \mathcal{N}}w_n^{*}H_n$ where $\mathcal{N}$ is the set of active Pauli operators and $w_n^{*}$ are the solutions to the linear program.
        Finally, the gradients of the state-wise distances can be derived (see Eq.~(\ref{eq:derivatives})), averaged and used to perform a gradient-based update of the generator \( G(\bm{\theta}) \). 
    
        \begin{figure}
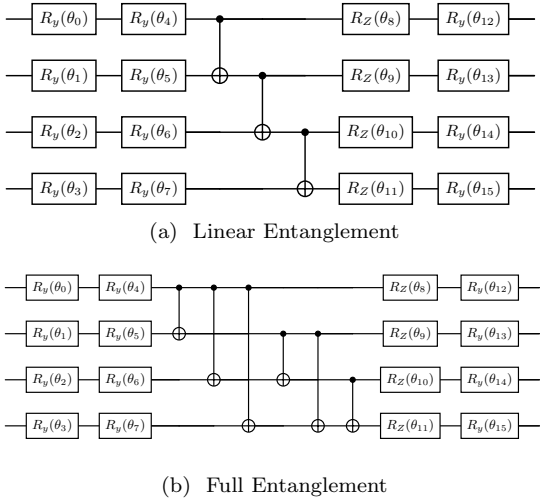

            \centering
            \subfloat[\label{subfig:HEA-linear} Linear Entanglement]{\subfile{circ-HEA-LE}}
            
            \subfloat[\label{subfig:HEA-full} Full Entanglement]{\subfile{circ-HEA-FE}}
    
            \caption{A single layer of hardware efficient ansatz (HEA) with $R_y$ and $R_z$ gates as rotation gates and two types of entanglement. (a) Linear entanglement where only nearest qubit is entangled (b) Full entanglement where every qubit is entangled to every other qubit}
            \label{fig:HEA}
        \end{figure}
        
\section{Experiments}
    \label{sec:exps}
    In this section, we will numerically evaluate QWC and benchmark it against HST and LET, focusing on susceptibility to barren plateaus. But before, we analyze the dependency on the $k$-locality of the discriminator and the size of the state ensemble needed for a successful compilation for different numbers of qubits.
    Since our primary goal is to show the viability of our chosen approach, we use the same circuit for the generator and the target. We fix the parameters of the target and randomly choose a different set of parameters for the ansatz. This ensures that at least one solution, i.e., set of parameters, exists for the compilation problem. 
    
    We specifically selected the hardware-efficient ansatz (HEA)~\cite{kandalaHardwareefficientVariationalQuantum2017} as our target and ansatz for demonstration. As large-scale implementations for chemistry~\cite{googleaiquantumandcollaborators*+HartreeFockSuperconductingQubit2020} and optimization~\cite{harriganQuantumApproximateOptimization2021} applications have shown, this ansatz leads to smaller errors due to hardware noise. The circuit diagram for a single layer HEA can be found in Fig.~\ref{fig:HEA}. 
    Additionally, we compare two distinct entanglement procedures to assess how the entangling property of the target unitary influences the required $k$-locality of the Pauli observables.

    In all experiments, we used the ADAM optimizer~\cite{DBLP:journals/corr/KingmaB14} with a learning rate of 0.1 for QWC and 0.04 for LET (HST) and exponential decay rates for the first and second moment estimates set as \(\beta_1 = 0.9\) and \(\beta_2 = 0.999\), respectively.

    \begin{figure}
        {\centering
        \subfloat[\label{subfig: k-locality_unsucc} Success percentage out of a total of 30 runs for different k-locality.]{      
            \includegraphics[width=0.4\textwidth]{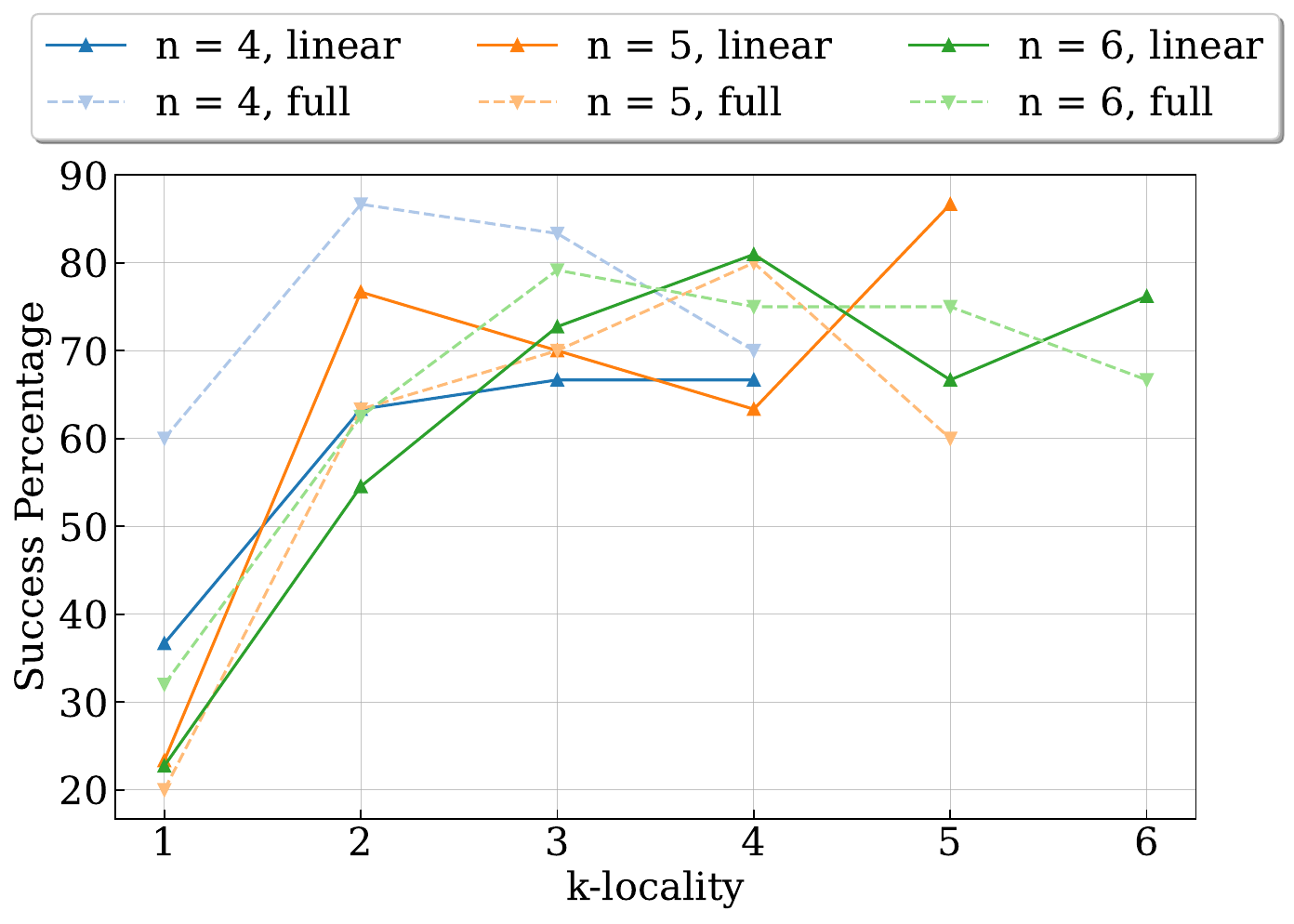}
        }
        \hfill
        \subfloat[\label{subfig: num_states_prob} Success percentage out of a total of 10 runs for different number of states used as input.]{ 
            \includegraphics[width=0.4\textwidth]{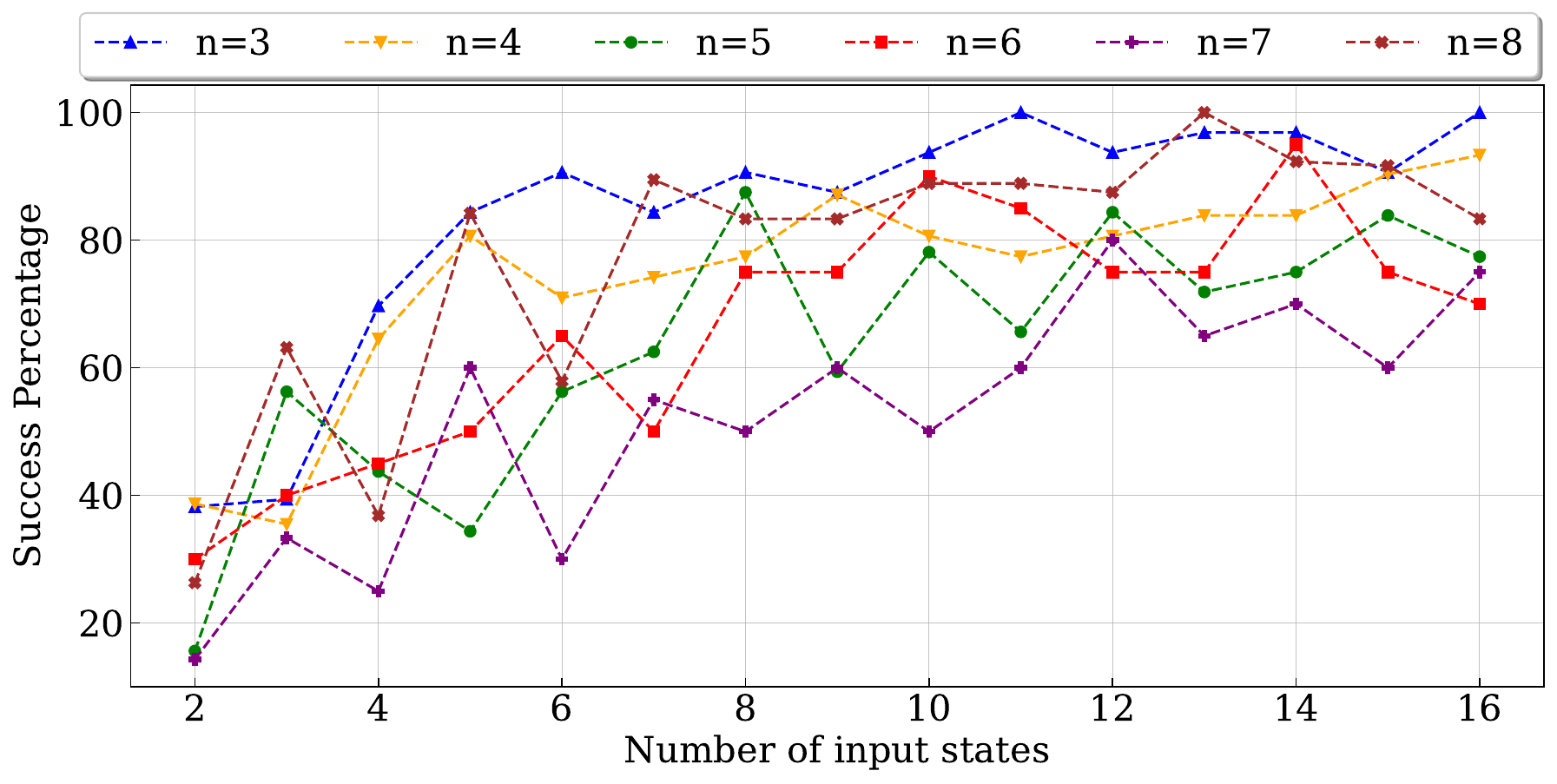}
        }
        \par}
        \caption{Experimental results for determining the $k$-locality and the amount of data (number of input states) required for successful compilation. (a) The number of $k$-local Pauli observables required to distinguish between the different types of entanglement. We take the 4-,5-, and 6-qubit single layer HEA with linear and full entanglement and run the compilation routine for each $k \in \{1,..,n\}$, where $n$ is the number of qubits under consideration, with 30 experiments each. The solid line shows the trend for linear entanglement, and the dashed line for full entanglement. (b) We fix $k = \lceil n/2 \rceil$ and use single layer HEA with linear entanglement. For successful compilation, the number of states which gives the highest success probability according to the plot, should be used.}
        \label{fig:data-demand}
    \end{figure}

    \subsection{Hyperparameters}
        \label{sec:hyperparams}
        Our compilation routine consists of the generator and the discriminator, each requiring hyperparameters related to the respective cost functions. We keep the target and the ansatz structure identical, in order to ensure guaranteed convergence, but the number of layers in the circuit is an important hyperparameter to see the effect of barren plateaus with increasing depth. Most of the hyperparameter search described below is carried out for a single-layer circuit. 
        
        We begin by defining successful compilation in terms of the cost function, whenever the cost function is below $10^{-3}$.
        In the previous section, we introduced the need for a test state ensemble for FUMC, i.e. a set $\probeset$ of quantum states that are used to calculate the empirical cost $\tilde{C}_{QW}^{(k)}(U,V,\probeset)$. The question then arises about the cardinality of this set and whether the set should be dynamically changed over the course of the training. 
        We found from our initial experiments that using a fixed set of states already gives successful training curves. This observation can also be interpreted as a test whether our set is large enough.
        For the discriminator, we mentioned that the expectation value of the Hamiltonian Eq.~(\ref{eq:hamiltonian}) needs to be evaluated for a $k$-local Pauli string. Here, $k$ is another hyper-parameter which needs to be tuned according to the problem. We show in Fig.~\ref{subfig: k-locality_unsucc} the success percentage over 30 experiments of compilation of a $4,5$ and $6$-qubit single layer HEA target ansatz pair, against the $k$-locality used to detect the entanglement in the target for two cases, linear and full entanglement. The two entangling circuits are shown in Fig.~\ref{fig:HEA}. We see a general trend of higher $k$ having higher success probability. Yet, a larger $k$ also translates to a higher number of observables. From observation, we choose to scale $k$ with $n$ as $k = \lceil n/2 \rceil$ for all following experiments.

    \subsection{Data Demand}
        \label{sec:data-demand}
        After choosing the $k$-locality for the discriminator and choosing a fixed state set $\probeset$, we conducted experiments to determine the number of states needed to achieve successful compilation.  For number of qubits $n \in \{3,...,8 \}$ we ran the training for $|\probeset| \in \{ 2,...,16 \}$ and calculated the fraction of runs which were successful out of a total of 10 runs for each state. We show the results in Fig.~\ref{subfig: num_states_prob}. We see the general trend that the success percentage increases as we increase the number of states used, which is what we expect. Yet, a higher number of states also requires higher computation time, and thus we must balance between successful compilation and amount of compute. For the rest of the experiments we chose the state set size $|\probeset| = 8$ for both QWC and LET.

        \begin{figure}[ht]
            \centering
            \subfloat[\label{subfig: hea-qubitwise} Function of number of qubits]{    \includegraphics[width=0.42\textwidth]{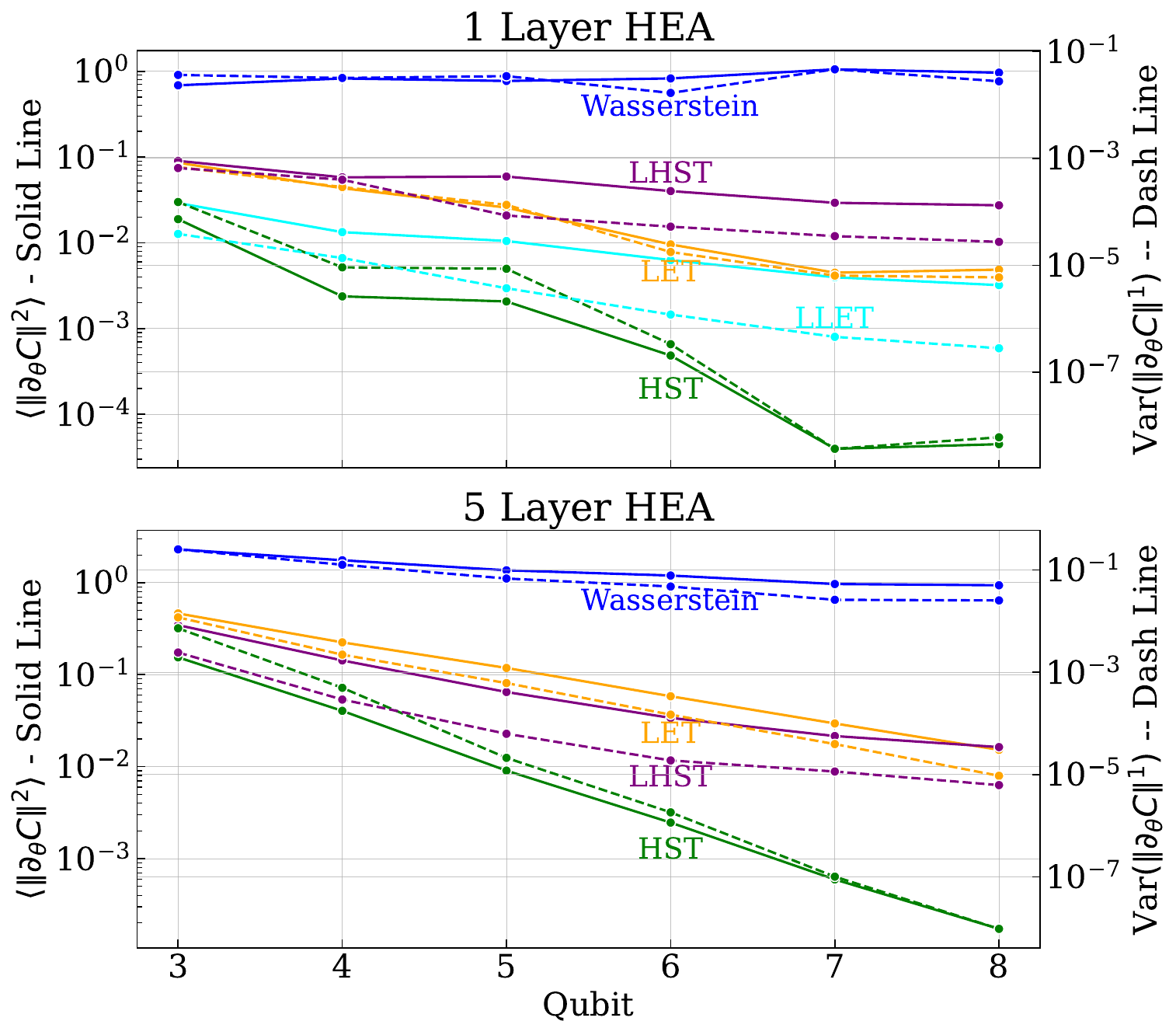}
            }
        
            \subfloat[\label{subfig: hea-layerwise} Function of number of layers]{ \includegraphics[width=0.42\textwidth]{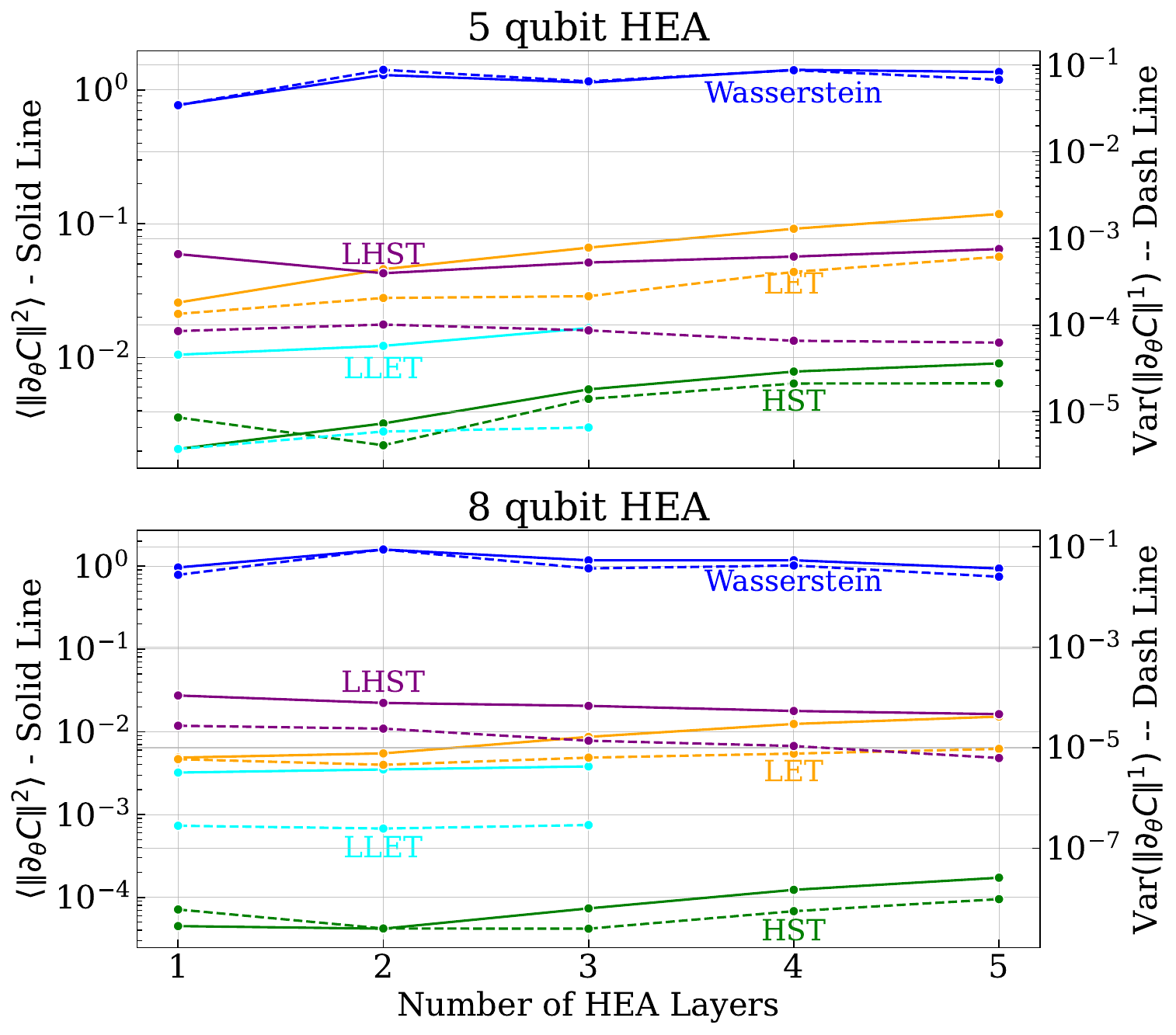}
            }
            
            \caption{Expectation and Variance of the $l_1$-norm of the gradient of the three cost functions, Wasserstein (our cost function), Hilbert-Schmidt test (HST), local HST, Loschmidt Echo test (LET) and local LET as a function of (a) number of qubits, (b) number of layers. The gradient is taken of the first parameter update step. Each point corresponds to the average over 100 runs.}
            \label{fig:barren_plateau}
        \end{figure}

    \subsection{Effects of Barren Plateaus}
        \label{sec:results-bp}
        To demonstrate that QWC is least affected by barren plateaus in the optimization landscape, we plot the expectation and variance of the $l_1$- norm of the gradient of the cost function with respect to the parameters of the ansatz as a function of (a) the number of qubits in the circuit and (b) the number of layers in the circuit. We consider a different number of layers~($1-5$) of the HEA for both the target and the ansatz. As before the number of layers is identical in both the target and ansatz. A single layer circuit is shown in Fig.~\ref{fig:HEA}(b). We follow the same approach as in Ref.~\cite{kianiLearningQuantumData2022} and calculate the gradients at the first optimization step. As before, we work with HEA as both target and ansatz, having full entanglement, restricting the Pauli observables set to $k = \lceil n/2 \rceil$-locality and $|\probeset|=8$ for all the qubits. The results are shown in Fig.~\ref{fig:barren_plateau}. We can see that the gradient norms of LET and HST decrease drastically as the number of qubits increases in both 1 layer and 5 layer circuits, indicating that these cost functions are adversely affected by the barren plateaus. For QWC, we see that for circuits with one layer and five layers, the gradient and the variance saturate as the number of qubits increases. As a function of the number of layers, there is no decay in the norms but the absolute values itself have a difference of orders of magnitude. Thus, we can conclude that QWC is least affected by barren plateaus compared to LET and HST. These results are consistent with the no-go theorems of Ref.~\cite{cerezoCostFunctionDependent2021}, since QWC uses local observables.

        \begin{figure}[ht]
            \subfloat[\label{subfig: 3q-hea-full} $n=3$]{    \includegraphics[width=0.46\textwidth]{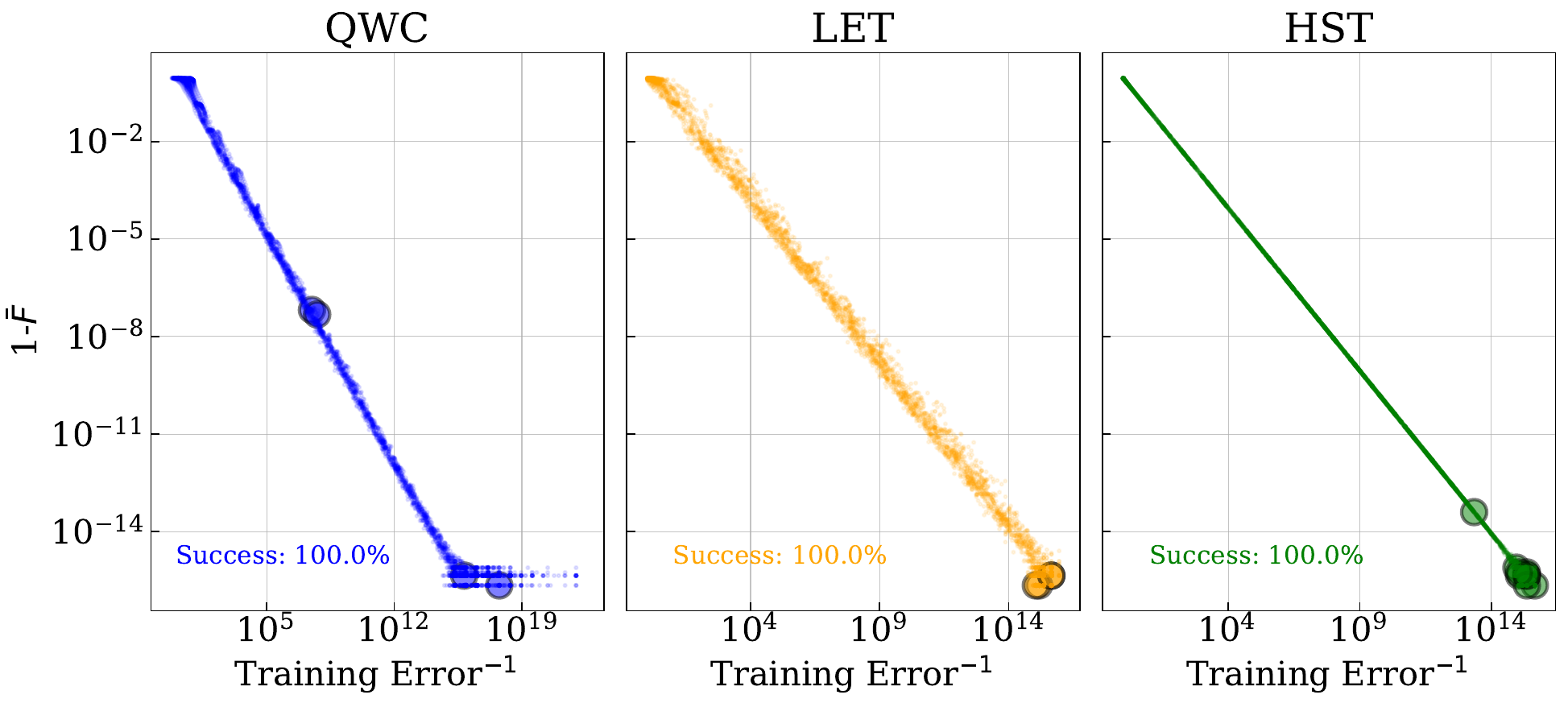}
            }
        
            \subfloat[\label{subfig: 4q-hea-full} $n=4$]{ \includegraphics[width=0.46\textwidth]{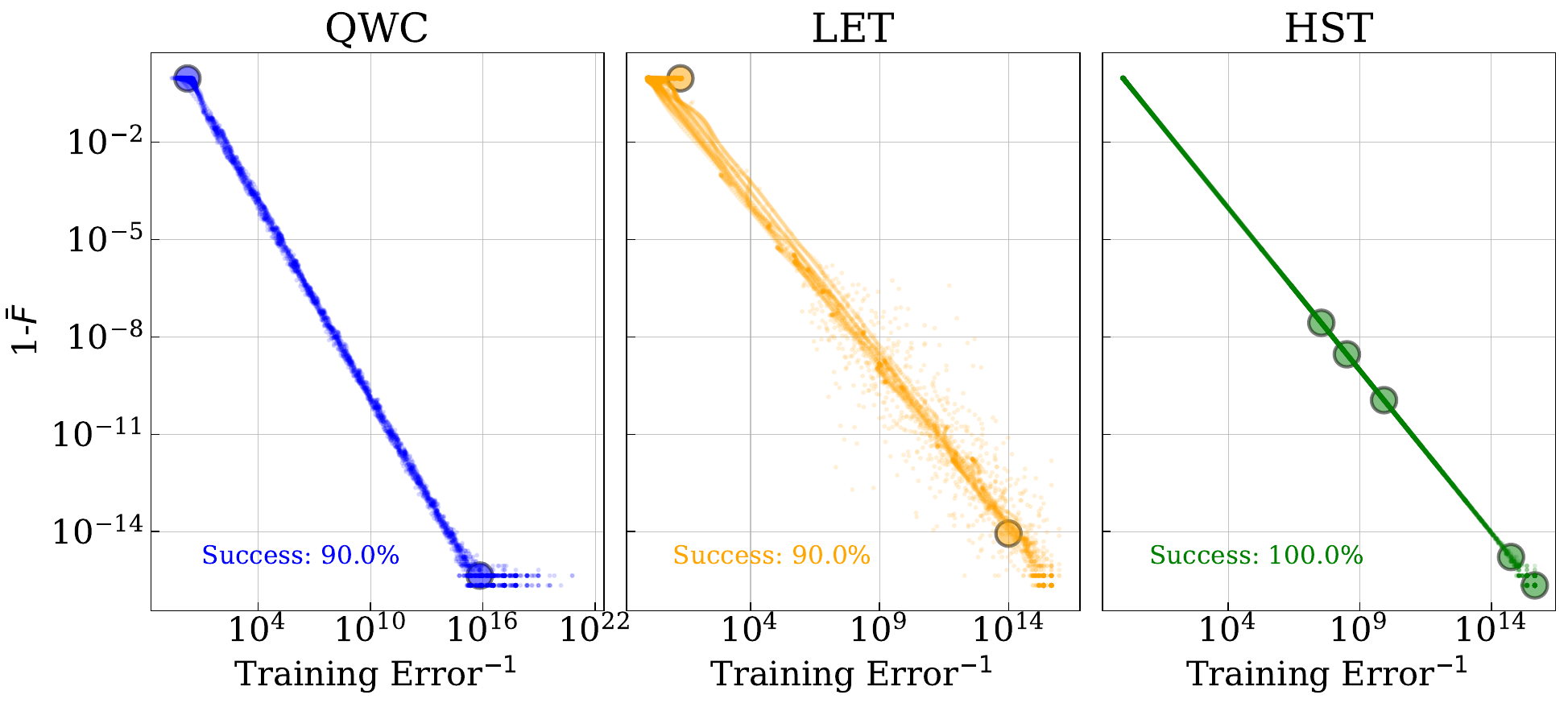}
            }
            \caption{Final infidelity ($1-\bar{F}$) vs. inverse training error ($C_{QW}^{-1})$ for hardware efficient ansatz (HEA) with full entanglement for $n=3$ and $n=4$ qubits. The training is carried out for 1000 steps. A run is successful when the cost function is below the threshold of $10^{-3}$. We see the trend that QWC like the other cost functions reaches low values of infidelity with a high probability.}
            \label{fig:inv-train-err-full}
        \end{figure}
        
    \subsection{Training results}
        \label{sec:results-training}
\begin{figure*}[th]
        \centering
        \subfloat[\label{subfig: 5q-hea-full}$n=5$]{%
                \includegraphics[width=0.5\textwidth]{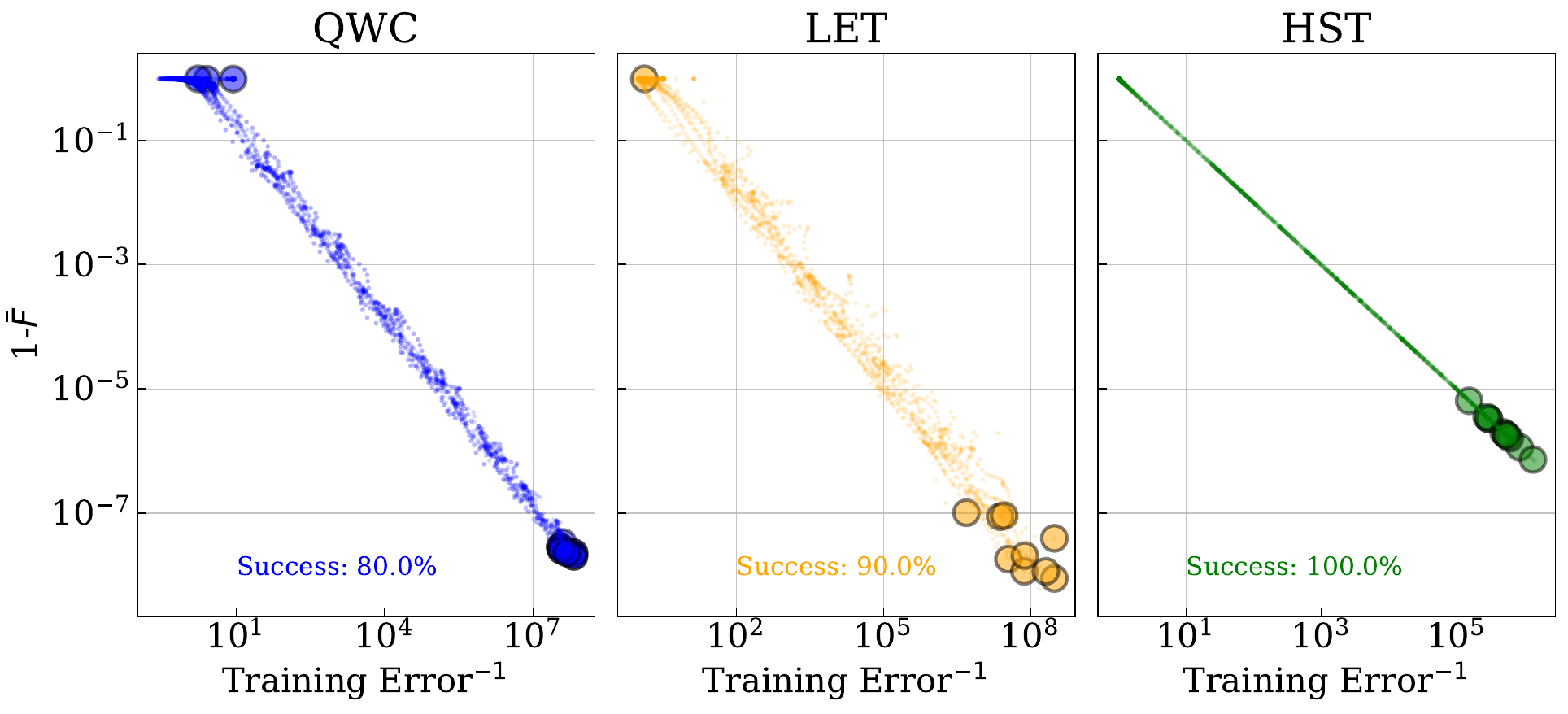}
        }
        \subfloat[\label{subfig: 6q-hea-full}$n=6$]{%
                \includegraphics[width=0.5\textwidth]{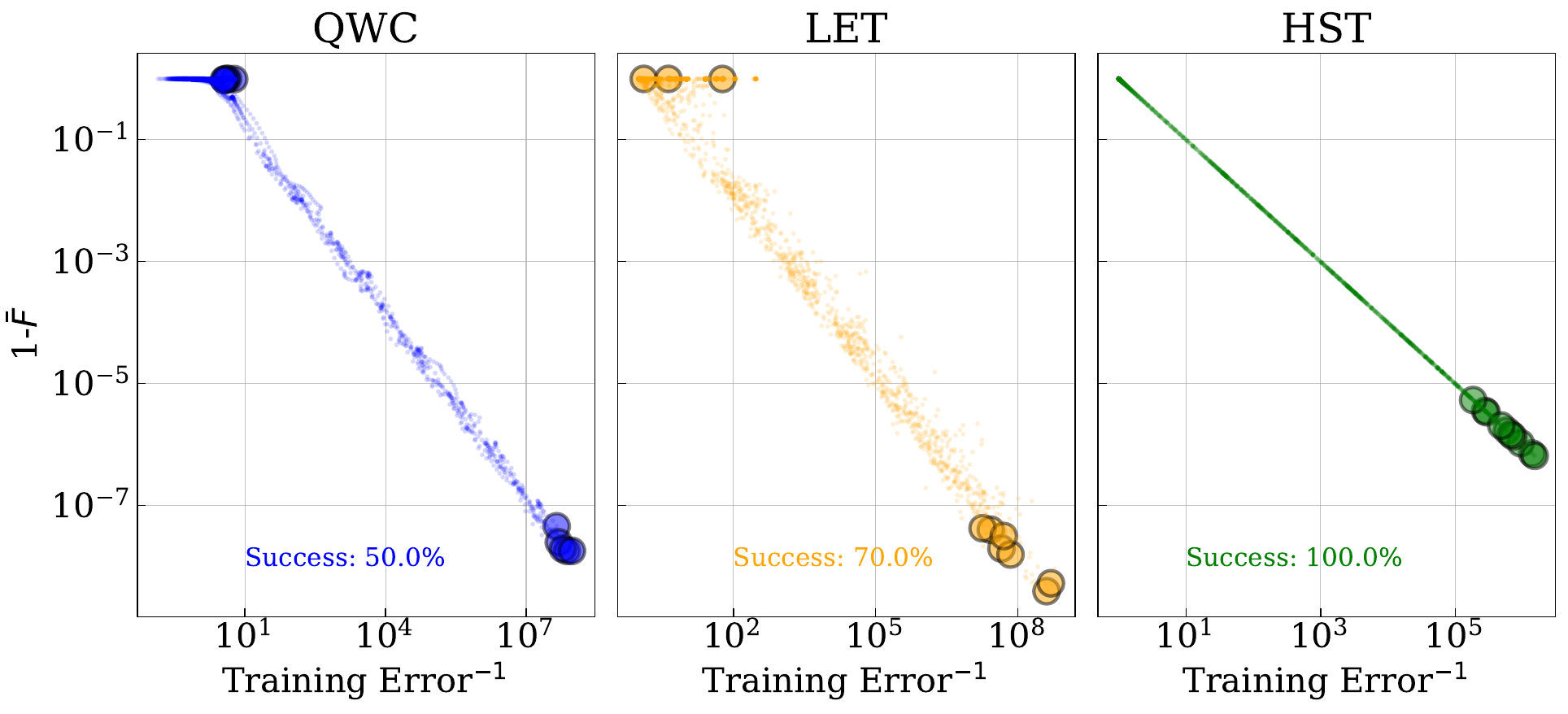}
        }
        
        \subfloat[\label{subfig: 7q-hea-full}$n=7$]{
          \includegraphics[width=0.5\textwidth]{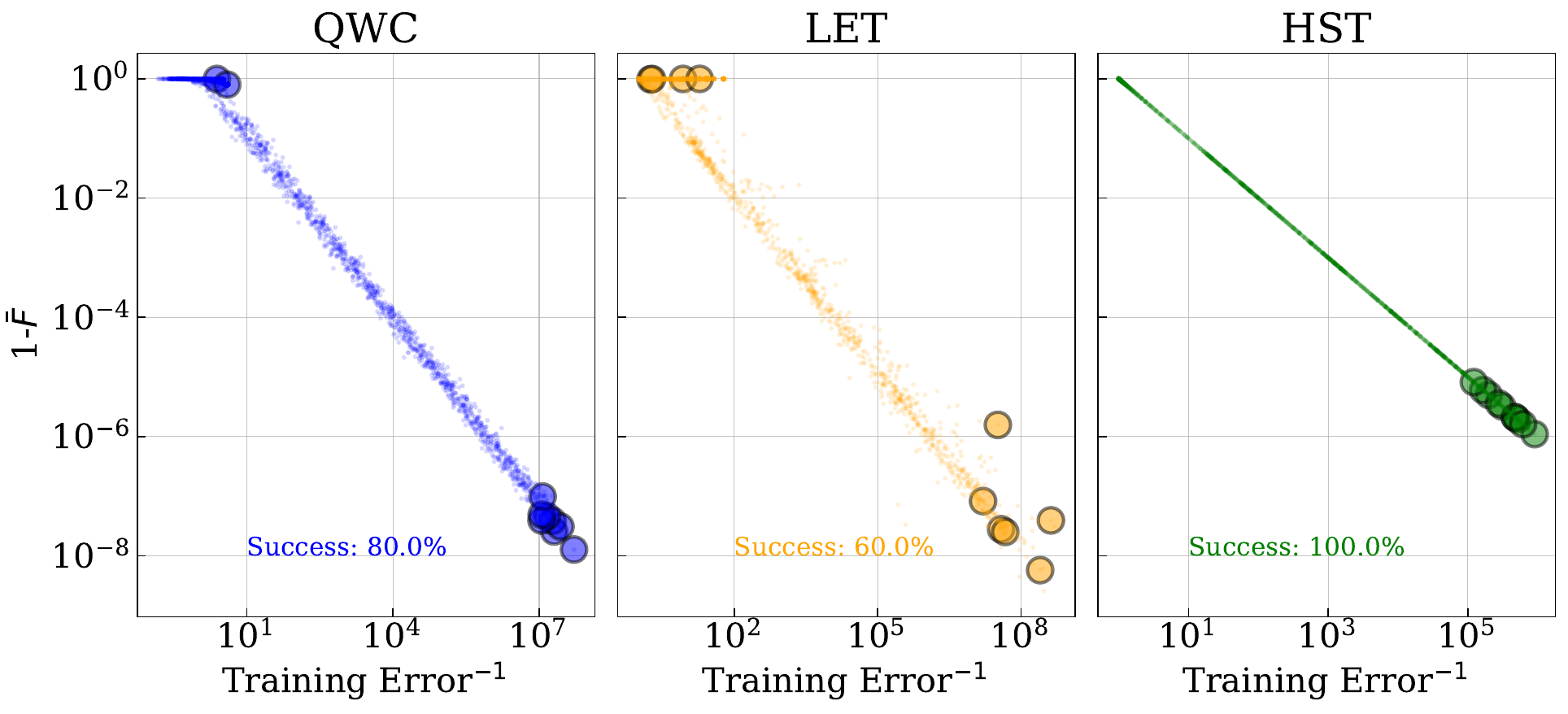}%
        }
        \subfloat[\label{subfig: 8q-hea-full}$n=8$]{%
          \includegraphics[width=0.5\textwidth]{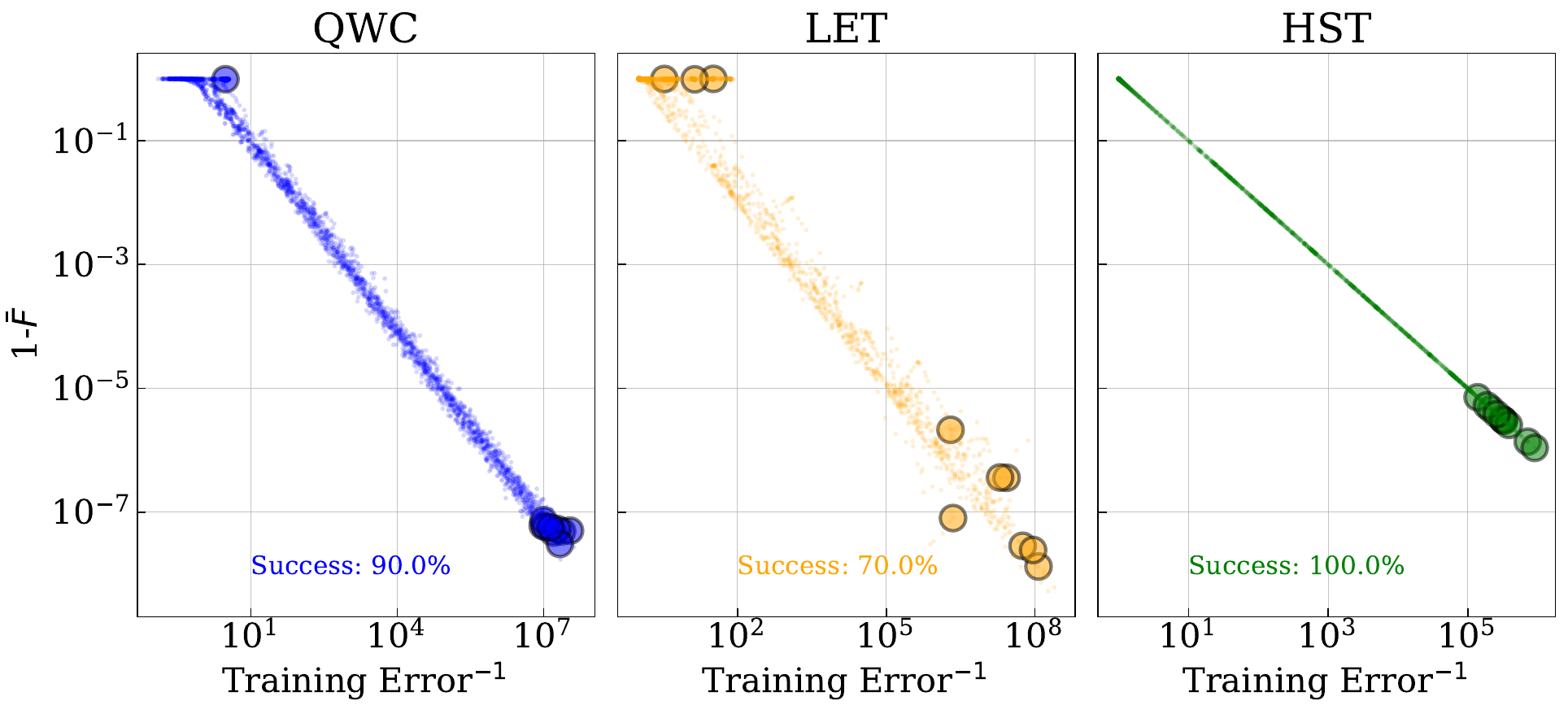}%
        }
        
            \caption{Final infidelity ($1-\bar{F}$) vs. inverse training error ($C_{QW}^{-1})$ for single layer HEA with full entanglement for $n \in \{ 5,..,8 \}$. Since most applications do not require infidelity values of order $10^{-15}$, here we employ early stopping of training when the variance of last 100 cost values reaches $10^{-8}$. }
            \label{fig:inv-train-err-earlystop}
        \end{figure*}

        The cost function Eq.~(\ref{eq:CEM-FBar}) is the metric we use to train our generator and discriminator, where when we reduce the cost $C_{QW}$ we are guaranteed that the infidelity between the test states also decreases, and the generator learns to mimic the target unitary. We show infidelity vs. inverse training error $C_{QW}^{-1}$ for the 3 and 4-qubit single-layer circuits in Figs.~\ref{subfig: 3q-hea-full} and \ref{subfig: 4q-hea-full}. We train for $1000$ steps and see that our cost function can reach infidelity values of $10^{-16}$, which is comparable to both LET and HST. Since such high precisions are usually not required in practical compilation routines, we plot in Fig.~\ref{fig:inv-train-err-earlystop} the same plots for $n \in \{5,..,8\}$ but with early-stopping. The early-stopping condition is invoked whenever the variance of the cost function in the last $100$ steps is less than $10^{-8}$. Both LET and HST reach convergence faster also with higher success rates compared to our method. In Fig.~\ref{fig:training} we plot the training curves for $n = 4, 6$ qubits to show convergence. Due to further hyper-parameter tuning, we do not plot the convergence results for multi-layered HEA structures.

    \subsection{Computation Details}
        \label{sec:comp-details}
        
        We make use of Qiskit v1.0~\cite{qiskit2024}, qiskit-aer v0.13.3, qiskit-algorithms v0.3 and qiskit-torch-module v0.1~\cite{meyer2024qiskit} with Python 3.10 for all our simulations. The hardware leverages AMD Ryzen Threadripper PRO 5965WX 24-Cores with 2 threads per core. The simulations make use of parallel processing of 8 cores by distributing the compilation for each of the~$|\probeset|$ states. As mentioned before, we make our implementation open-source in the GitHub repository~\cite{Richter_quantum-wasserstein-compilation_2025}.
        
\section{Conclusion}
    \label{sec:conclusion}

    We have introduced a novel cost function for variational quantum circuit compilation, based on the Wasserstein distance of order 1 which has the property of not being unitarily invariant. Our approach can leverage quantum computers to estimate circuit similarity through a framework that combines aspects of both quantum state discrimination and generative adversarial networks.
    We proved that this QWC cost function provides an upper bound for the average infidelity between unitary transformations, establishing its theoretical validity for circuit compilation.
    
    Through numerical experiments, we demonstrated that the one-step gradients of our cost function are least affected by the presence of barren plateaus as we scale to larger qubit numbers and deeper circuits. Further numerical simulations on circuits with 3 to 8 qubits (single-layer HEA) revealed several important insights. The effectiveness of the discriminator strongly depends on the locality of available Pauli observables, with insufficient locality leading to overestimated similarities. Although our method requires more measurements (scaling as $\mathcal{O}(n^k)$) compared to traditional approaches, it showed a clear correlation between infidelity and compilation cost when given sufficient locality. We also demonstrated that compilation can be achieved effectively using simultaneous measurements on a fixed set of randomly sampled test states. However, the optimal training data requirements remain an open question.
    
    A comparative analysis revealed that while HST achieved better success rates, it becomes impractical for larger systems due to its requirement for twice the number of qubits. The primary limitation of QWC is the scaling of measurement observables as the qubit count increases. However, recent research on classical estimation techniques~\cite{angrisaniClassicallyEstimatingObservables2024, mangini2024low} suggests potential improvements in this area. 
    Furthermore, we did not conduct experiments on deeper circuits because they require extensive hyperparameter tuning. We believe that there will be no increase in the number of Pauli observables needed, compared to the shallow experiments, and only a slight increase in the number of states required for successful compilation, is expected.

    Furthermore, classical estimation techniques can be easily integrated into our framework, which could accelerate the training process. As of now, our results indicate that QWC does not provide immediate advantages over HST or LET. However, once we integrate the classical estimation techniques into our framework, we anticipate significant performance improvements in both time and scaling. 
    Lastly, while our current study focused on noiseless simulations, exploring noise resilience, similar to the work done for HST and LET in Ref.~\cite{sharmaNoiseResilienceVariational2020}—represents an important direction for future research.

\section*{Acknowledgement}
The research is part of the Munich Quantum Valley (MQV) and was supported by the Bavarian Ministry of Economic Affairs, Regional Development and Energy with funds from the Hightech Agenda Bayern via the project BayQS.

\bibliographystyle{quantum}
\bibliography{references}

\appendix

\section{Quantum \texorpdfstring{$W_1$}{W\_1} distance and Fidelity}
    \label{sec:em-to-fidelity}
    As explained in Section ~\ref{sec:compilation}, the standard measure of success in variational quantum compilation is the average fidelity $\Fbar(U,V)$, Eq. (\ref{eq:haar-av-fid}). Naturally, the question arises: what is the relation between the average quantum $W_1$ distance $C_{QW}(U,V)$ (Eq.~\ref{eq:CEM-exact}) and $\Fbar(U,V)$?

    The starting point for our derivation is Proposition 2 of \cite[]{depalmaQuantumWassersteinDistance2021} that states upper and lower bounds for the quantum $W_1$ norm in terms of the trace norm $\norm{\cdot}_1$.
    \begin{align}
        \label{eq:W1-norm-bounded-Trace-norm}
        \frac{1}{2}\norm{\rho - \sigma}_1 \leq \Wnorm{\rho - \sigma } \leq \frac{n}{2} \norm{\rho - \sigma}_1 \;.
    \end{align}
    Additionally, the trace norm is bounded by $F(\rho, \sigma)$:
    \begin{align}
        1 - \sqrt{F(\rho, \sigma)} \leq \frac{1}{2} \norm{\rho - \sigma}_1 \leq \sqrt{1 - F(\rho, \sigma)} \;.
    \end{align}
    Hence, we can find a lower bound for the fidelity in terms of the quantum $W_1$ norm:
    \begin{align}
        1 -  \Wnorm{\rho-\sigma} \leq \sqrt{F(\rho, \sigma)} \;.
    \end{align}
    Since the fidelity is bounded, $0 \leq F(\rho, \sigma) \:\forall\: \rho, \sigma \in \mathcal{S(\hilb)}$, the same holds for $\sqrt{F(\rho, \sigma)}$. We will now constrain the quantum $W_1$ norm to small values, $0 \leq \Wnorm{\rho-\sigma} \leq 1$. This domain is of particular interest as we formulate the VQC problem as a minimization of the quantum $W_1$ norm. With this constraint, we can square the inequality and make use of Bernoulli's inequality:
    \begin{align}
        F(\rho, \sigma) \geq \left(1 - \Wnorm{\rho-\sigma}\right)^2 \geq 1 - 2 \Wnorm{\rho - \sigma} \;.
    \end{align}
    By this bound, we now know that a vanishing Earth Mover's distance between two mixed states translates to high fidelity of the states. But this result for mixed states only holds for small distances, e.g. $\Wnorm{\rho - \sigma} \leq 1$.
    
    Since QWC actually uses pure states, a more general result can be found for this case. For two pure states $\rho=\ketbra{\psi}, \sigma=\ketbra{\phi}$, the following equality between trace norm and fidelity $F(\ket{\psi}, \ket{\phi})=\abs{\braket{\psi|\phi}}^2$ holds:
    \begin{align}
        \big\lVert\ketbra{\psi}-\ketbra{\phi}\big\rVert_1 = \sqrt{1-F(\ket{\psi}, \ket{\phi})} \;.
    \end{align}
    Using again Eq.~(\ref{eq:W1-norm-bounded-Trace-norm}), we bound the fidelity by the quantum $W_1$ norm,
    \begin{align}
        \big\lVert \ketbra{\psi}-\ketbra{\phi}\big\rVert_{W_1} \geq  \sqrt{1-F(\ket{\psi},\ket{\phi})} \;,
    \end{align}
    and square without further constraints:
    \begin{align}
        \big\lVert\ketbra{\psi}-\ketbra{\phi}\big\lVert_{W_1}^2 \geq  1-F(\ket{\psi},\ket{\phi}) \;.
    \end{align}
    This upper bound for the infidelity of pure states in terms of the quantum $W_1$ norm motivates Def.~\ref{def:qwc} as the squared $W_1$ distance. 
    \vspace{0.5cm}
\section{Gradients of the Empirical Cost Function}
    \newcommand{\bempCEM}[1]{\widetilde{C}_\text{EM}\big(#1\big)}
    
    \label{sec:qwc-gradients}
    In Section~\ref{sec:qwc-empirical}, we define the cost function to estimate the restricted quantum EM distance (Eq.~\ref{eq:empirical_CEM}). Since we focus on gradient-based optimization, we need to provide the derivative of the cost function $\bempCEM{U(t),V,\probeset}$, here written for a single parameter $t$ representing a parameter in the parameterized ansatz~$U$.
    \begin{proposition}
        \label{prop:gradients}
        Let $V$ be a unitary operator on $\hilb$ and $U(t)$ a parametric family of unitary transformations on $\hilb$. Then, the derivative of the empirical Wasserstein compilation cost in parameter $t$ can be expressed as
        \begin{align}
            \label{eq:gradients}
            \begin{split}
                \diff*{\bempCEM{U(t),V, \probeset}}{t}[t=0] & \\ =  \sum_{\psi\in\probeset} \frac{2}{|\probeset|} W_1\Bigl(U(0) \ket{\psi}, V \ket{\psi}\Bigr)\cdot & \\ \cdot W_1'\Bigl(U(0) \ket{\psi}, V \ket{\psi}\Bigr) \;,
            \end{split}
        \end{align}
        where $\probeset$ is a state ensemble and $W_1'$ can be calculated according to Eq.~(49) of Ref. \cite{kianiLearningQuantumData2022}.
    \end{proposition}
    
    \begin{figure*}[t]
        \centering
        \subfloat[\label{subfig: GlobHST} HST]{\begin{tikzpicture}
  \begin{yquant*}[register/separation=2mm, every nobit output/.style={},
  /yquant/operator/minimum width=2mm]
    {
    \yquantset{every multi label/.style={every node/.style={anchor=east, midway}}}
    init {$\ket{0_n}_{A}$} (a[-2]);
    }
    {
        \yquantset{every multi label/.style={every node/.style={anchor=east, midway}}}
        init {$\ket{0_n}_{B}$}(b[-2]);
    }

      hspace {2mm} -;

      [name=Estart]
      h a;
      cnot b[0] | a[0];
      cnot b[1] | a[1];
      [name=Eend]
      cnot b[2] | a[2];

      hspace {1mm} -;
      [x radius=.25cm]
      box {$U$} (a);
      hspace {2mm} -;
      [x radius=.25cm]
      box {$V^\dagger$} (a);
      hspace {1mm} -;

      [name=EdaggerStart]
      cnot b[2] | a[2];
      cnot b[1] | a[1];
      cnot b[0] | a[0];
      [name=EdaggerEnd]
      h a;

      hspace {2mm} -;

      measure a,b;

  \end{yquant*}
  \node[draw,dashed, fit=(Estart-0) (Eend), "\small$E$"] {};
  \node[draw,dashed, fit=(EdaggerStart) (EdaggerEnd-0), "\small$E^\dagger$"] {};
\end{tikzpicture}}  \hfill
        \subfloat[\label{subfig: LHST} LHST]{\begin{tikzpicture}
    \begin{yquant*}[register/separation=1.0mm, every nobit output/.style={}]
       {
        \yquantset{every multi label/.style={every node/.style={anchor=east, midway}}}
        init {$\ket{0^n}_{A}$} (a[-2]);
        }
        {
            \yquantset{every multi label/.style={every node/.style={anchor=east, midway}}}
            init {$\ket{0^n}_{B}$}(b[-2]);
        }

        hspace {1mm} -;
        [this subcircuit box style={dashed, "$E$"}]
        subcircuit {
            qubit {} a[3];
            qubit {} b[3];
            h a;
            cnot b[0] | a[0];
            cnot b[1] | a[1];
            cnot b[2] | a[2];
            }
            (a,b);

        hspace {.5mm} -;

        [x radius=.3cm]
        box {$U$} (a);
        hspace {1.5mm} -;
        [x radius=.3cm]
        box {$V^\dagger$} (a);

        hspace {1mm} -;
        [this subcircuit box style={dashed, "$(E^{(1)})^\dagger$"}]
        subcircuit {
            qubit {} a[3];
            qubit {} b[3];

            hspace {0mm} -;
            cnot b[0] | a[0];

            h a[0];
            }
            (a,b);

        hspace {1mm} -;

        measure a[0],b[0];

    \end{yquant*}
\end{tikzpicture}}
    
        \subfloat[\label{subfig:GlobLET} LET]{\begin{tikzpicture}
    \begin{yquant*}[register/separation=4mm, every nobit output/.style={}]
       {
        \yquantset{every multi label/.style={every node/.style={anchor=east, midway}}}
        init {$\ket{0_n}$} (a[-2]);
        }

        hspace {1mm} -;
        [x radius=.3cm]
        box {$W$} (a);
        [thick, label=$\ket{\psi_0} $]
        barrier (a);

        [x radius=.3cm]
        box {$U$} (a);
        hspace {2mm} -;
        [x radius=.3cm]
        box {$V^\dagger$} (a);

        hspace {2mm} -;
        [x radius=.3cm]
        box {$W^\dagger$} (a);

        hspace {2mm} -;

        measure a;

    \end{yquant*}
\end{tikzpicture}} \hspace{10mm}
        \subfloat[\label{subfig:LLET} LLET]{\begin{tikzpicture}
    \begin{yquant*}[register/separation=5mm, every nobit output/.style={}]
       {
        \yquantset{every multi label/.style={every node/.style={anchor=east, midway}}}
        init {$\ket{0_n}$} (a[-2]);
        }

        hspace {1mm} -;
        [x radius=.3cm]
        box {$W$} (a);
        [thick, label=$\ket{\psi_0} $]
        barrier (a);

        [x radius=.3cm]
        box {$U$} (a);
        hspace {2mm} -;
        [x radius=.3cm]
        box {$V^\dagger$} (a);

        hspace {2mm} -;
        [x radius=.3cm]
        box {$W^\dagger$} (a);

        hspace {2mm} -;

        measure a[0];

    \end{yquant*}
\end{tikzpicture}}
        \caption{Quantum circuits of metrics for FUMC. The circuits are reproduced from \cite{sharmaNoiseResilienceVariational2020}. (a) The probability of the all-zero outcome is equivalent to the Hilbert-Schmidt inner product $\abs{\Trace{(V^\dagger U)}}^2/d^2$. Maximizing this probability compiles $V$ into the target unitary $U$ (see Eq.~(\ref{eq:HST})). (b) The local Hilbert-Schmidt test is an adaptation for higher qubit numbers. The cost function is built from the mean of the pairwise 00 probabilities. (c) In the Loschmidt Echo test, the initial state is prepared using the $W$ unitary and the overlap is measured with the unitarily evolved $V^{\dagger}U$ state by measuring for the all zero-state on all qubits. (d)~The local LET is used for higher qubits number, by taking the mean of single qubit measurements.}
        \label{fig:qc-HST}
    \end{figure*}

    \begin{proof}
        The proof follows by simply applying the sum rule and the chain rule for derivatives on the definition of the empirical cost function:
        \begin{widetext}
            \begin{eqnarray}
                \diff*{\bempCEM{U(t),V, \probeset}}{t}[t=0] & = &  \diff*{\frac{1}{|\probeset|} \sum_{\psi\in\probeset} W_1^2\Bigl(U(t) \ket{\psi}, V \ket{\psi}\Bigr)}{t}[t=0]    =  \frac{1}{|\probeset|} \sum_{\psi\in\probeset} \diff*{ W_1^2\Bigl(U(t) \ket{\psi}, V \ket{\psi}\Bigr)}{t}[t=0]   \\ \nonumber
                & = & \sum_{\psi\in\probeset} \frac{2}{|\probeset|} W_1\Bigl(U(0) \ket{\psi}, V \ket{\psi}\Bigr) \diff*{W_1\Bigl(U(t) \ket{\psi}, V \ket{\psi}\Bigr)}{t}[t=0] \;.
            \end{eqnarray}
        \end{widetext}
    
    \end{proof}

    From Prop. \ref{prop:gradients}, we can see that acquiring the gradient requires estimating the $W_1$ distance once for each state and, additionally, twice per parameter and per state for the derivative $\difs{W_1}{t}$ if we use standard techniques like the parameter-shift rule \cite{schuldEvaluatingAnalyticGradients2018}.

\section{Cost Functions for Variational Compilation}
    \label{sec:cost-functions-for-vqcc}
    In this appendix, we are giving details on the other VQCC cost functions applied in our numerical simulations, namely HST and LET.
    
    We show the quantum circuit for the Hilbert-Schmidt test in Fig.~\ref{subfig: GlobHST}. The cost function $\CHST$ is faithful, i.e. vanishes if and only if $U=V$ (up to a global phase), and has by Eq.~(\ref{eq:HST-Fbar_relatn}) an operational meaning~\cite{khatriQuantumassistedQuantumCompiling2019}. 
    To address the issue of barren plateaus~\cite{cerezoCostFunctionDependent2021}, the local Hilbert-Schmidt (LHST) test was introduced~\cite{khatriQuantumassistedQuantumCompiling2019}. LHST is a local adoption of HST where the entanglement fidelities~$F_{\text{LHST}}^{(j)}$ of local quantum channels between the~$j$-th qubit of each subsystem are measured:
    
    \begin{align}
        \label{eq:lhst-cost}
        \mathcal{C}_{\text{LHST}} = 1 - \frac{1}{n} \sum_{j=1}^{n} F_{\text{LHST}}^{(j)} \;.
    \end{align}

    Another cost function in VQCC is based on the idea of the Loschmidt echo \cite{goussevLoschmidtEcho2012}. Governed by a Hamiltonian~$H_1$, the forward evolution by time ~$t$ is followed by the application of a second Hamiltonian~$-H_2$ to recover the initial state~$\ket{\psi_0}$, defining the Loschmidt echo as 
    \begin{align}
        \label{eq:loschmidt}
        M(t)=\abs{\braket{\psi_0 | e^{iH_2t/\hbar}e^{-iH_1t/\hbar} | \psi_0}}^2.
    \end{align}
    It quantifies the recovery of an initial quantum state after the application of an imperfect time-reversal procedure \cite{goussevLoschmidtEcho2012}. It is directly accessible by the circuit drawn in Fig.~\ref{subfig:GlobLET} called the Loschmidt echo test. Here, $W$ is used to prepare the input-state different from $\ket{0_n}$. The cost function $\CLET$ suffers from the same scaling issues as $\CHST$ since it applies a global cost function. A possible resolution to this problem was again suggested in terms of local measurements, and the quantum circuit for the same is shown in Fig.~\ref{subfig:LLET}.
    
    \section{Additional Plots of Training}
\begin{figure*}
            \centering
            \subfloat[\label{subfig:4q-HEA-full-training} $n=4$]{
                \includegraphics[width=0.95\textwidth]{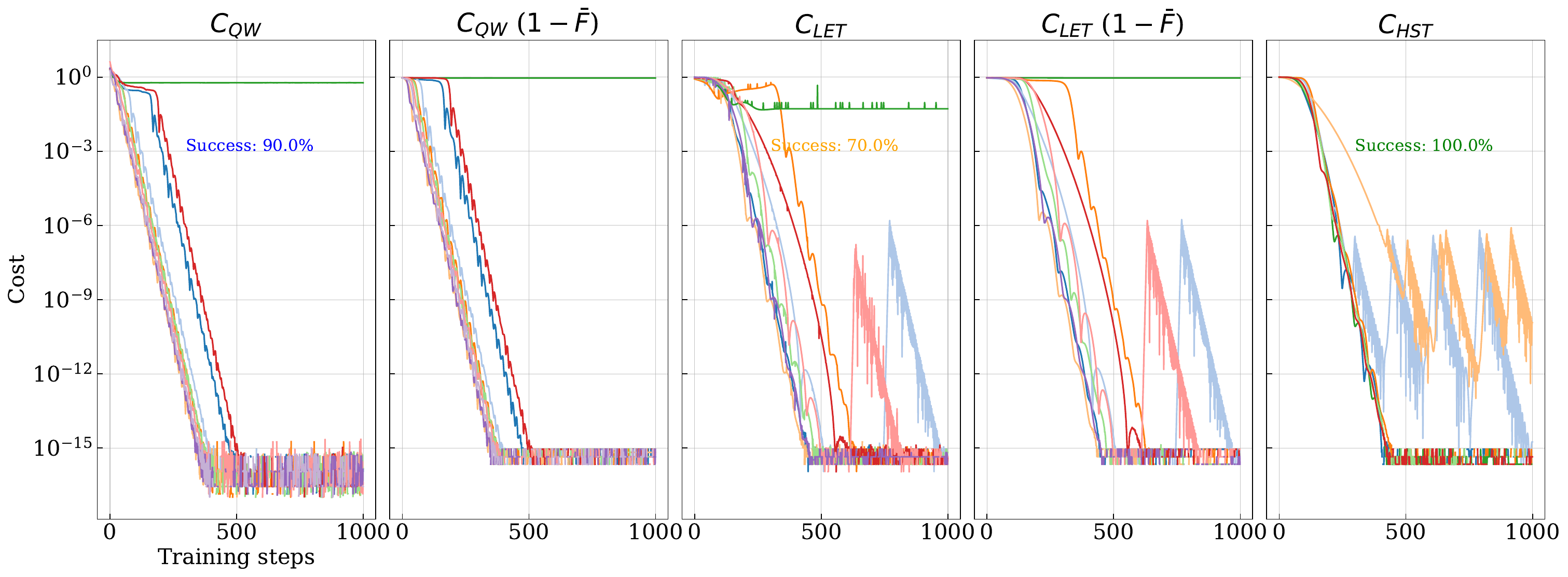}
            }
            \vspace{1cm}
            \subfloat[\label{subfig:6q-HEA-full-training} $n=6$]{
                \includegraphics[width=0.95\textwidth]{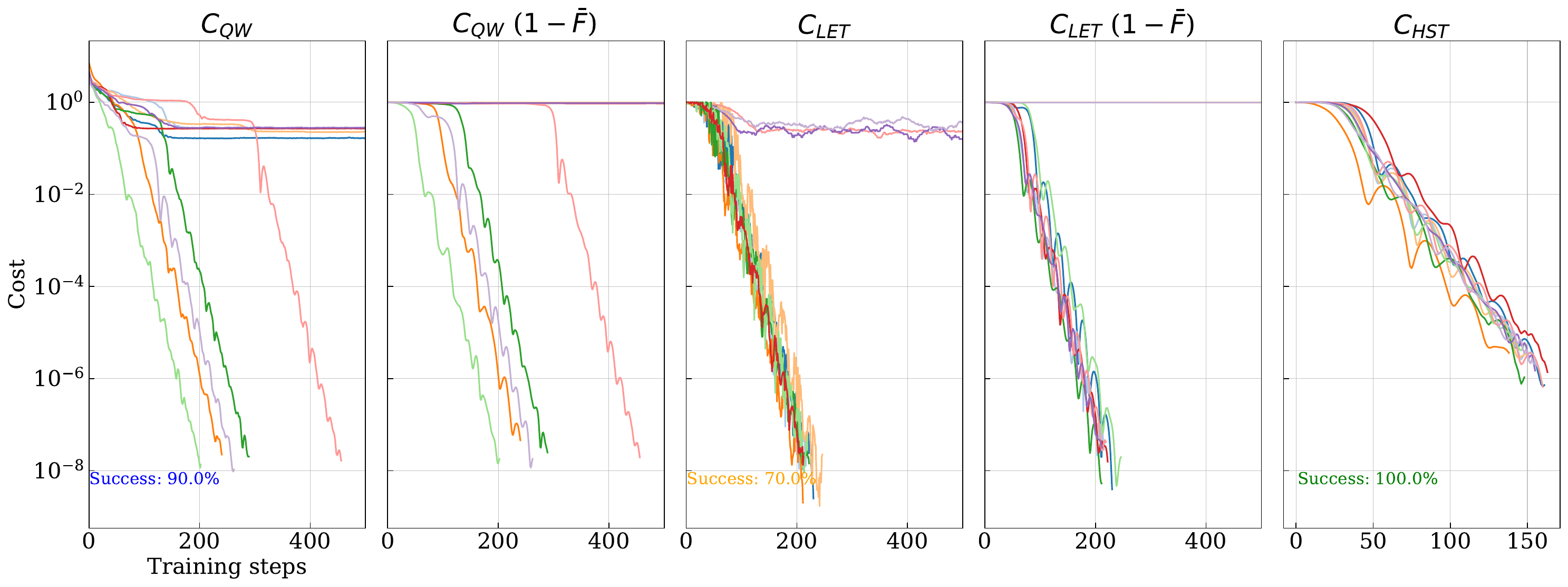}
            }
            \caption{Training curves for the 4-qubit and 6-qubit target ansatz pair for HEA with full entanglement. (a) The training is continued for the full 1000 steps in order to verify if all the methods reach the same global optimum. (b) Here, early stopping is employed, where the training is stopped if the last 100 values of the variance of the cost function reaches $10^{-8}$. We see that in this case LET and HST reach convergence faster than QWC.}

            \label{fig:training}

        \end{figure*}

    This appendix provides with Fig. \ref{fig:training} more visualizations of typical training dynamics for different cost function approaches used in our VQCC experiments. The visible spikes for LET and HST in Fig. \ref{subfig:4q-HEA-full-training} are numerical instabilities due to small numbers. This can be avoided by employing early-stopping as shown in Fig. \ref{subfig:6q-HEA-full-training}.

\end{document}